\documentclass[twocolumn]{aastex63}

\usepackage{graphicx}
\usepackage[T1]{fontenc}
\usepackage{aecompl}
\usepackage[]{amsmath}

\usepackage{color}
\usepackage{float}

\usepackage{xspace}
\usepackage{soul}
\usepackage{placeins}
\usepackage{atbegshi}
\newcommand{\handlethispage}{}
\newcommand{\discardpagesfromhere}{\let\handlethispage\AtBeginShipoutDiscard}
\newcommand{\keeppagesfromhere}{\let\handlethispage\relax}
\newcommand{\msun}{\mathrm{M}_\odot}
\graphicspath{{figures/}}

\def\lsim{ \lower .75ex \hbox{$\sim$} \llap{\raise .27ex \hbox{$<$}} }

\submitjournal{ApJ}

\shorttitle{Velocity anisotropy for halo stars, dark matter and subhalos}
\shortauthors{He et al.}
\graphicspath{{./}{figures/}}
\begin{document}

\title{How does the velocity anisotropy of halo stars, dark matter and satellite galaxies depend on host halo properties?}

\correspondingauthor{Wenting Wang}
\email{wenting.wang@sjtu.edu.cn}

%%\author[0000-0002-0786-7307]{Greg J. Schwarz}
%%\affiliation{American Astronomical Society \\
%%667 K Street NW, Suite 800 \\
%%Washington, DC 20006, USA}

\author{Jiaxin He}
\affiliation{Department of Astronomy, Shanghai Jiao Tong University, Shanghai 200240, China}
\affiliation{Shanghai Key Laboratory for Particle Physics and Cosmology, Shanghai 200240, China}
\author[0000-0002-5762-7571]{Wenting Wang}
\affiliation{Department of Astronomy, Shanghai Jiao Tong University, Shanghai 200240, China}
\affiliation{Shanghai Key Laboratory for Particle Physics and Cosmology, Shanghai 200240, China}
\author[0000-0001-7890-4964]{Zhaozhou Li}
\affiliation{Centre for Astrophysics and Planetary Science, Racah Institute of Physics, The Hebrew University, Jerusalem, 91904, Israel}
\author[0000-0002-5762-7571]{Jiaxin Han}
\affiliation{Department of Astronomy, Shanghai Jiao Tong University, Shanghai 200240, China}
\affiliation{Shanghai Key Laboratory for Particle Physics and Cosmology, Shanghai 200240, China}
\author{Vicente Rodriguez-Gomez}
\affiliation{Instituto de Radioastronom\'ia y Astrof\'isica, Universidad Nacional Aut\'onoma de M\'exico, Apdo. Postal 72-3, 58089 Morelia, Mexico}
\author{Donghai Zhao}
\affiliation{Shanghai Astronomical Observatory, Chinese Academy of Sciences, 80 Nandan Road, Shanghai 200030, China}
\author{Xianguang Meng}
\affiliation{Shanghai Astronomical Observatory, Chinese Academy of Sciences, 80 Nandan Road, Shanghai 200030, China}
\author[0000-0002-4534-3125]{Yipeng Jing}
\affiliation{Department of Astronomy, Shanghai Jiao Tong University, Shanghai 200240, China}
\affiliation{Shanghai Key Laboratory for Particle Physics and Cosmology, Shanghai 200240, China}
\author{Shi Shao}
\affiliation{Key Laboratory for Computational Astrophysics, National Astronomical Observatories, Chinese Academy of Sciences, Beijing 100101, China}
\author{Rui Shi}
\affiliation{Department of Astronomy, Shanghai Jiao Tong University, Shanghai 200240, China}
\affiliation{Shanghai Key Laboratory for Particle Physics and Cosmology, Shanghai 200240, China}
\author{Zhenlin Tan}
\affiliation{Department of Astronomy, Shanghai Jiao Tong University, Shanghai 200240, China}
\affiliation{Shanghai Key Laboratory for Particle Physics and Cosmology, Shanghai 200240, China}

%\collaboration{1}{(AAS Journals Data Scientists collaboration)}



\begin{abstract}

We investigate the mass ($M_{200}$) and concentration ($c_{200}$) dependencies of the velocity anisotropy ($\beta$) profiles for different components in the dark matter halo, including halo stars, dark matter  and subhalos, using systems from the IllustrisTNG simulations. Beyond a critical radius, $\beta$ becomes more radial with the increase of $M_{200}$, reflecting more prominent radial accretion around massive halos. The critical radius is $r\sim r_s$, $0.3~r_s$ and $r_s$ for halo stars, dark matter and subhalos, with $r_s$ the scale radius of host halos. This dependence on $M_{200}$ is the strongest for subhalos, and the weakest for halo stars. In central regions, $\beta$ of halo stars and dark matter particles gets more isotropic with the increase of $M_{200}$ in TNG300 due to baryons. By contrast, $\beta$ of dark matter from the dark matter only TNG300-Dark run shows much weaker dependence on $M_{200}$ within $r_s$. Dark matter in TNG300 is slightly more isotropic than in TNG300-Dark at $0.2~r_s<r<10~r_s$ and $\log_{10}M_{200}/\msun<13.8$. Halo stars and dark matter also become more radial with the increase in $c_{200}$, at fixed $M_{200}$. Halo stars are more radial than the $\beta$ profile of dark matter by approximately a constant beyond $r_s$. Dark matter particles are more radial than subhalos. The differences can be understood as subhalos on more radial orbits are easier to get stripped, contributing more stars and dark matter to the diffuse components. We provide a fitting formula to the difference between the $\beta$ of halo stars and of dark matter at $r>r_s$ as $\beta_\mathrm{star}-\beta_\mathrm{DM}=(-0.028\pm 0.008)\log_{10}M_{200}/\msun + (0.690\pm0.010)$.

%\revised{proper motion errors or photometric errors, or both?}
%PM errors, thank you for pointing out! 

%When $M(200-300\mathrm{pc})$ is under-estimated, we are likely to end up with more core-ish best fits, when 
%the truth is more cuspy.

\end{abstract}



\section{Introduction}
\label{sec:intro}

The velocity anisotropy, which is defined as one minus the ratio between tangential and radial velocity dispersions for a population of objects ($\beta=1-\frac{\sigma^2_t}{2\sigma^2_r}$), describes the dominance or fraction of radial and tangential orbits for the population. It is a concise parameter commonly measured and investigated for a variety of tracers in galaxy kinematics and Milky Way (MW) science. 

On one hand, the velocity anisotropy is frequently used to study the dynamical status and orbital properties for observed galaxy systems, hence probing their formation and evolution. For example, it is well recognized that early type galaxies are composed of isotropic stellar orbits \citep[e.g.][]{2014MNRAS.442.3284A}. Early type galaxies themselves in galaxy cluster systems follow nearly isotropic orbits, whereas late type galaxies follow slightly more radial orbits \citep[e.g.][]{2004A&A...424..779B}. Redshift evolution in the orbital anisotropy of cluster galaxies was also reported
%, which is likely due to the mass growth of galaxy clusters during their fast accretion phase 
\citep{2009A&A...501..419B}. The orbital properties for smaller galaxies upon infalling into larger systems strongly affect the subsequent evolution of the internal features of galaxies \citep{2012MNRAS.427.1024I}. The velocity anisotropy parameter is also adopted to study the orbits of global clusters (GC) in and around galaxies, and it was shown that metal rich GCs are tangentially anisotropic while metal poor GCs tend to be more isotropic or radial. The tangential features of metal rich GCs in the inner region could originate from the depletion of radial orbits by tidal stripping, and accretion could make the orbits more radial in the outer part \citep[e.g.][] {2014MNRAS.442.3299A,2016MNRAS.462.4001Z}.
%\cite{2021MNRAS.500.3151S}

For our MW halo stars, the measurements of their velocity anisotropies suffer from relatively larger inconsistencies before the {\it Gaia} era \citep[e.g.][]{1998AJ....115..168C,2004AJ....127..914S,2012MNRAS.424L..44D,2012ApJ...761...98K,2014ApJ...794...59K,2015ApJ...813...89K,2017ApJ...846...10A,2016MNRAS.463.2623H,2017ApJ...846...10A}. Many of the studies report dips in the profiles, likely associated with substructures. The robustness of the measurements depends on a variety of factors, such as selection effects and observational errors. After large amount of proper motion data collected by \textit{Gaia} and line-of-sight velocities collected by LAMOST, \cite{2019AJ....157..104B} measured the velocity anisotropy for a large sample of K giants. Between the solar radius and 25~kpc to the Galactic center, their sample is highly radial, and it gradually drops beyond 25~kpc, reaching $\sim$0.3 at $\sim$100~kpc. \cite{2019AJ....157..104B} did not report any dip for the velocity anisotropy profile within 25~kpc. They claimed the sensitivity of their measurements to substructures \cite[also see][]{2018ApJ...853..196L}. In follow-up studies, the velocity anisotropy of MW halo stars is reported to be less radial with decreasing metallicity \citep[e.g.][]{2021ApJ...919...66B}, and also differs between different types of stars and depends on sky fields \citep[e.g.][]{2019ApJ...879..120C}, which carry important information about the assembly history of our host Galaxy. There also exist many studies looking into the anisotropy of halo stars in MW-like systems from simulations, with comparisons to the real MW \citep[e.g.][]{2022ApJ...937...20E,2022JCAP...10..058P}.

On the other hand, precise determination of the velocity anisotropy for the tracer population is important for proper constraints on the underlying dark matter distribution for our MW Galaxy, nearby dwarf galaxies and galaxy clusters. Individual stars can be resolved in our MW and nearby dwarfs, and they can be used as tracer objects to infer the underlying dark matter distribution according to their kinematics (see \cite{2020SCPMA..6309801W} for a review). For galaxy clusters, line-of-sight velocities of companion satellite galaxies, and sometimes resolved globular clusters or planetary nebula in nearby systems, can be used as tracers. However, the velocity anisotropies for observed tracers have to be known in order to break the mass-anisotropy degeneracy and properly constrain the rotation velocity or the underlying potential. Unfortunately, if tangential velocities are not available or with large errors, one has to rely on extra assumptions to model the anisotropy parameter\citep[e.g.][]{2005MNRAS.364..433B,2009ApJ...704.1274W,2009MNRAS.395...76D,2012ApJ...761...98K,2013MNRAS.429.3079M,2014ApJ...794...59K}, hence preventing precise constraints of the dark matter distribution. 

Given the importance of using the velocity anisotropy parameter to describe the dynamical status and orbital properties of tracer populations, and in constraining the underlying dark matter potential model of the host system, the purpose of this paper is to use high resolution hydrodynamical simulations to investigate how do the velocity anisotropies of different halo/galaxy components (accreted halo stars, dark matter and subhalos/satellites) depend on the properties of the host dark matter halos. Proper understanding of how the velocity anisotropy of tracer objects depends on their host halo mass can potentially help us setting up priors upon, for example, constraining our MW halo mass. 

Many previous studies have adopted numerical simulations to investigate how does the velocity anisotropy of dark matter particles depend on the host halo properties. For example, a universal relation between the slopes of dark matter density profiles and the velocity anisotropy parameter of dark matter particles were reported \citep{2006NewA...11..333H}, which can help to deduce a simple estimator from the Jeans Equation to infer the rotation velocities \citep{2010ApJ...718L..68H}. On average, the velocity anisotropy for dark matter particles in numerical simulations is close to isotropic in the very central regions, which gradually increase with the increase in radius, and eventually drop beyond the virial radius, where the motions are dominated by more coherent radial infall with decreased radial dispersions\citep[e.g.][]{2005MNRAS.361L...1W,2008MNRAS.386.2022A,2012ApJ...752..141L}.  
There are large system to system scatters, with the differences depending on many factors such as the halo mass, redshift and relaxation status of the systems. The velocity anisotropy is also reported to have a strong dependence on direction, likely due to the merger histories, which 
may explain the large scatter in the outer part of halos \citep{2012JCAP...07..042S}. A detailed investigation on how the velocity anisotropy of dark matter particles in numerical simulations depends on host halo properties, density peak, formation histories and relaxation status will also be presented in Meng et al. (in preparation). Compared with dark matter particles in numerical simulations, the velocity anisotropy profiles of subhalos in host halos are found to be similar, but lower than that of dark matter in the inner region \citep{2004MNRAS.352..535D}.
  
In this paper, we are not only going to look at the velocity dispersion of dark matter particles, but we will also focus more on the velocity anisotropy for halo stars and subhalos, and compare the difference between the velocity anisotropies of different components. The stellar and dark halos all grow through accretion, and thus we expect they would share similarities regarding how their velocity anisotropies depend on the host halo properties. We achieve our goal by using the modern Illustris TNG suite of hydrodynamical simulations.  

The layout of this paper is as follows. We introduce our sample of halo/galaxy systems selected from the Illustris TNG series of simulations in Section~\ref{sec:data}. Our method of calculating the velocity anisotropy for accreted halo star particles, dark matter particles and subhalos is introduced in Section~\ref{sec:method}. Results will be presented in Section~\ref{sec:results}, and we conclude in the end (Section~\ref{sec:concl}). 

\section{Data}
\label{sec:data}

\subsection{TNG simulations}

In this study, we adopt the IllustrisTNG suites of simulations. Sophisticated baryonic processes are incorporated with the moving-mesh code \citep[\textsc{arepo};][]{Springel2010}, including, for example, metal line cooling, star formation and evolution, chemical enrichment and gas recycling. The TNG simulations adopt the Planck 2015 $\Lambda$CDM cosmological model with $\Omega_\mathrm{m}=0.3089$, $\Omega_\Lambda=0.6911$, $\Omega_\mathrm{b}=0.0486$, $\sigma_8=0.8159$, $n_s=0.9667$, and $h=0.6774$ \citep{Planck2015}. A total of 100 snapshots are saved between redshifts of $z=20$ and $z=0$, with the initial condition set up at $z=127$. For more details, we refer the readers to \cite{Marinacci2018,Naiman2018,Nelson2018,Pillepich2018,Springel2018} and \cite{Nelson2019}. 

We mainly use the TNG300-1 simulation in our analysis, which has the largest box size and thus the best statistics. In the end of this paper we will also show comparisons with TNG100-1 and TNG50-1, that enable us going down to smaller mass ranges due to their higher resolutions than TNG300-1. TNG300-1, TNG100-1 and TNG50-1 are the simulations with the highest resolution in the corresponding suite (compared with TNG50/100/300-2 and TNG50/100/300-3), and hereafter we call them TNG300, TNG100 and TNG50 for short. The box sizes of TNG300, TNG100 and TNG50 are 205~Mpc/h, 75~Mpc/h and 35~Mpc/h on a side. These numbers are approximately 50, 100 and 300~Mpc, so this is the reason for their name conventions. 

TNG300 follows the joint evolution of 2,500$^3$ dark matter particles and $\sim$2,500$^3$ gas cells. Each dark matter particle has a mass of $4.0\times10^7 \msun$/h, while the baryonic mass resolution is $7.6\times10^6 \msun$/h. TNG100 follows the joint evolution of 1,820$^3$ dark matter particles and approximately 1,820$^3$ gas cells. Each dark matter particle has a mass of $5.1\times10^6 \msun$/h, while the baryonic mass resolution is $9.4\times10^5 \msun$/h. TNG50 follows the joint evolution of 2,160$^3$ dark matter particles and approximately 2,160$^3$ gas cells. Each dark matter particle has a mass of $3.1\times10^5 \msun$/h, while the baryonic mass resolution is $5.7\times10^4 \msun$/h.

In addition to the series of hydrodynamical TNG runs introduced above, we will use the corresponding dark matter only (DMO) version of TNG300, which we denote as TNG300-Dark. Comparing the velocity anisotropy of dark matter particles in TNG300 and TNG300-Dark, we can gain intuitions on the effect of baryons.

\subsection{Our sample of galaxy systems}

We select only central galaxies from TNG simulations. In this paper, we will divide central galaxies into different mass bins according to the virial mass\footnote{We adopt the convention for virial mass, denoted as $M_{200}$, as the total mass enclosed within the virial radius, $R_{200}$, of the host dark matter halos, with $R_{200}$ defined as the radius within which the total matter density is 200 times the critical density of the universe.}, $M_{200}$, of their host dark matter halos. Limited by the mass resolution of the original simulations, we should set up a lower limit in the host halo mass, below which the stellar halos might not be realistic. We determine this lower mass limit in $M_{200}$ for different resolutions of TNG simulations according to the following logic. Starting from the average mass of star particles in TNG300, TNG100 and TNG50, we require that a realistic satellite galaxy should have at least 100 star particles, which leads to the lower limit in the stellar mass of satellites ($M_{\ast,\mathrm{sat,limit}}$). Halo stars are stripped from bound satellites to form the stellar halos of central galaxies. To have realistic stellar halos and velocity anisotropy measurements of these halo stars, we require that the stellar mass of central galaxies should be at least one order of magnitude higher than $M_{\ast,\mathrm{sat,limit}}$. We then transfer this lower limit in the stellar mass of central galaxies to the lower limit in the host halo virial mass, $M_{200}$, according to the corresponding stellar mass to halo mass relation in the simulation, and we take the upper bound in $M_{200}$ at fixed stellar mass. According to this reasoning, we choose the lowest mass that we can reach as $\log_{10}M_{200}/\msun\sim 12.6$ for TNG300, $\log_{10}M_{200}/\msun\sim 11.8$ for TNG100 and $\log_{10}M_{200}/\msun\sim 11.4$ for TNG50.

When calculating the velocity anisotropy for accreted halo stars, we only use ex-situ formed star particles from the simulations. Ex-situ stars are defined as those star particles that are stripped from accreted smaller satellite galaxies, while in-situ stars are formed by gas cooling. Explicitly, we use the stellar assembly catalogue provided by the TNG website \citep{2016MNRAS.458.2371R,Rodriguez-Gomez2017} to identify the origin of star particles. Throughout this paper, our definition of halo stars in TNG simulations are those ex-situ star particles. We avoid the usage of in-situ star particles, that are not formed through accretion and may complicate our understandings. 

To calculate the velocity anisotropy of dark matter, we simply use all bound dark matter particles of each system, but with particles belonging to satellites/subhalos removed. For subhalos, we use those that have at least 20 dark matter particles and are bound to the system within the Friends-of-Friends (FoF) group.

\begin{figure}
\centering
\includegraphics[width=1\linewidth]{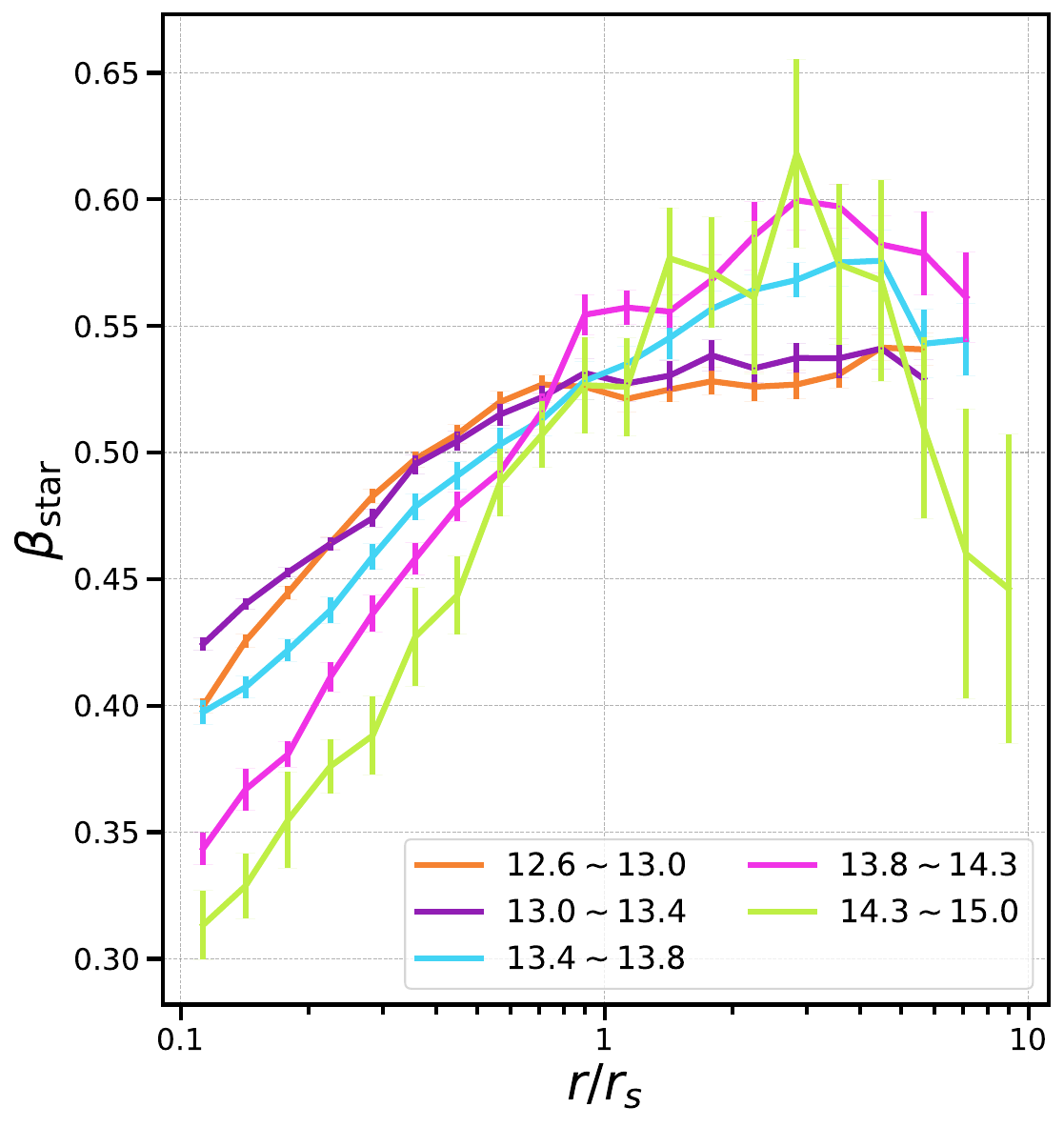}
\caption{The velocity anisotropy for halo stars, reported as a function of the galactocentric radius, $r$, divided by the NFW scale radius, $r_s$. We first calculate the velocity anisotropy profiles for particles in each individual galaxy system, and each curve refers to the median of a population of galaxies in the same bin of $M_{200}$. The legend shows the corresponding range in $\log_{10}M_{200}/\msun$. The errorbars are 1-$\sigma$ scatters of 100 bootstrapped subsamples of galaxies in the same mass bin.}
\label{fig:beta_star_rs}
\end{figure}

\section{Method}
\label{sec:method}

The velocity anisotropy is often denoted by the symbol $\beta$. For a population of particles, it is defined through Equation~\ref{eqn:beta} below
\begin{equation}
\beta=1-\frac{\sigma^2_\theta+\sigma^2_\phi}{2\sigma^2_r}
=1-\frac{\langle {v^2_\theta} \rangle- \langle v_\theta \rangle^2+\langle {v^2_\phi}\rangle-\langle v_\phi \rangle^2}{2(\langle {v^2_r} \rangle-\langle v_r \rangle^2)}, 
\label{eqn:beta}
\end{equation}
where $\sigma_r$ is the radial velocity dispersion. $\sigma_\theta$ and $\sigma_\phi$ are tangential velocity dispersions for the two tangential velocity components. $v_r$ is the radial velocity. $v_\theta$ and $v_\phi$ are the tangential velocities. $\beta$ can go from minus infinity to 1. When radial orbits dominate the population, $\beta$ is close to 1, and when tangential orbits dominate, $\beta$ is negative. $\beta=0$ means the orbits are isotropic. In this study, we will calculate the velocity anisotropy for halo stars, dark matter particles and subhalos from TNG simulations. 

To calculate $\beta$ as a function of galactocentric radius for objects in galaxy systems binned according to their halo properties from TNG simulations, {\it we first calculate $\beta$ for particles or objects in each individual galaxy system, and then for galaxies in the same bin of halo mass or concentration, we report the median. This is the convention adopted for all measurements and figures presented in this paper.} Here we do not mix particles from different galaxy systems together to calculate $\beta$, to avoid massive systems having a large number of particles to dominate the result. 

Note, however, for a given galaxy system, sometimes the number of particles in a given galactocentric radius bin is too few to give a proper estimate of the velocity dispersions and $\beta$, and thus for a given bin in galactocentric radius, we include a criterion that only galaxy systems with more than 10 star or dark matter particles can contribute to the final median estimate of this bin, when calculating the $\beta$ profiles for halo stars and dark matter. When calculating the $\beta$ profile for subhalos, because the number of subhalos is significantly smaller than those of star and dark matter particles, we decrease this threshold to 5 subhalos in each radial bin. For halo stars, dark matter particles and subhalos, we all require that the fraction of galaxy systems excluded due to this criterion should be less than 20\%, otherwise we will not plot results for this bin. %For subhalos, this would mainly eliminate measurements at small radius, because the number of subhalos is too few in the center. For star particles, this would mainly eliminate measurements at large radius, because the number density profile of the halo stars drops very fast. 

\begin{figure}
\centering
\includegraphics[width=1\linewidth]{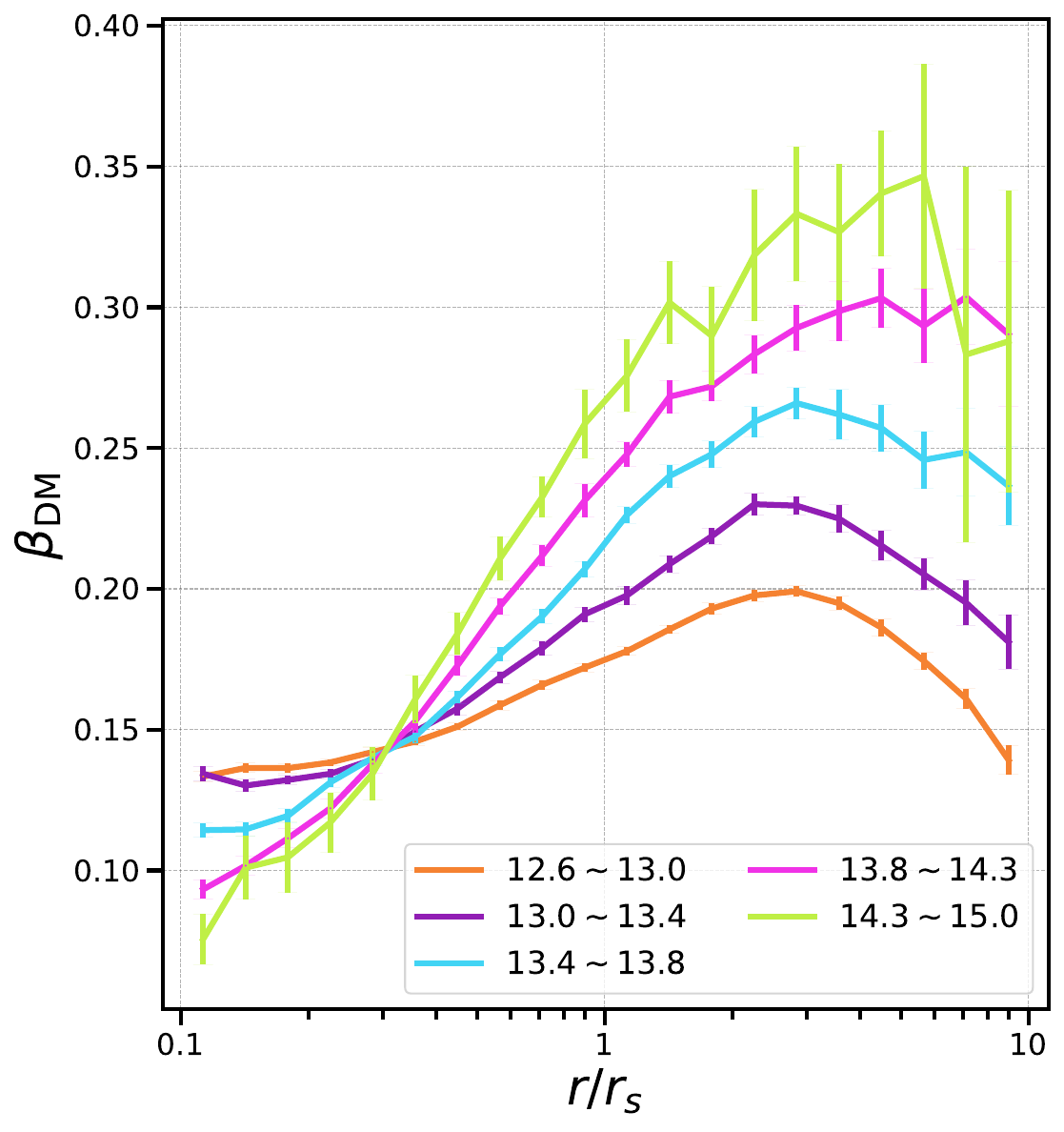}
\caption{Similar to Figure~\ref{fig:beta_star_rs}, but for dark matter particles from TNG300. }
\label{fig:beta_dm_rs}
\end{figure}

\begin{figure}
\centering
\includegraphics[width=1\linewidth]{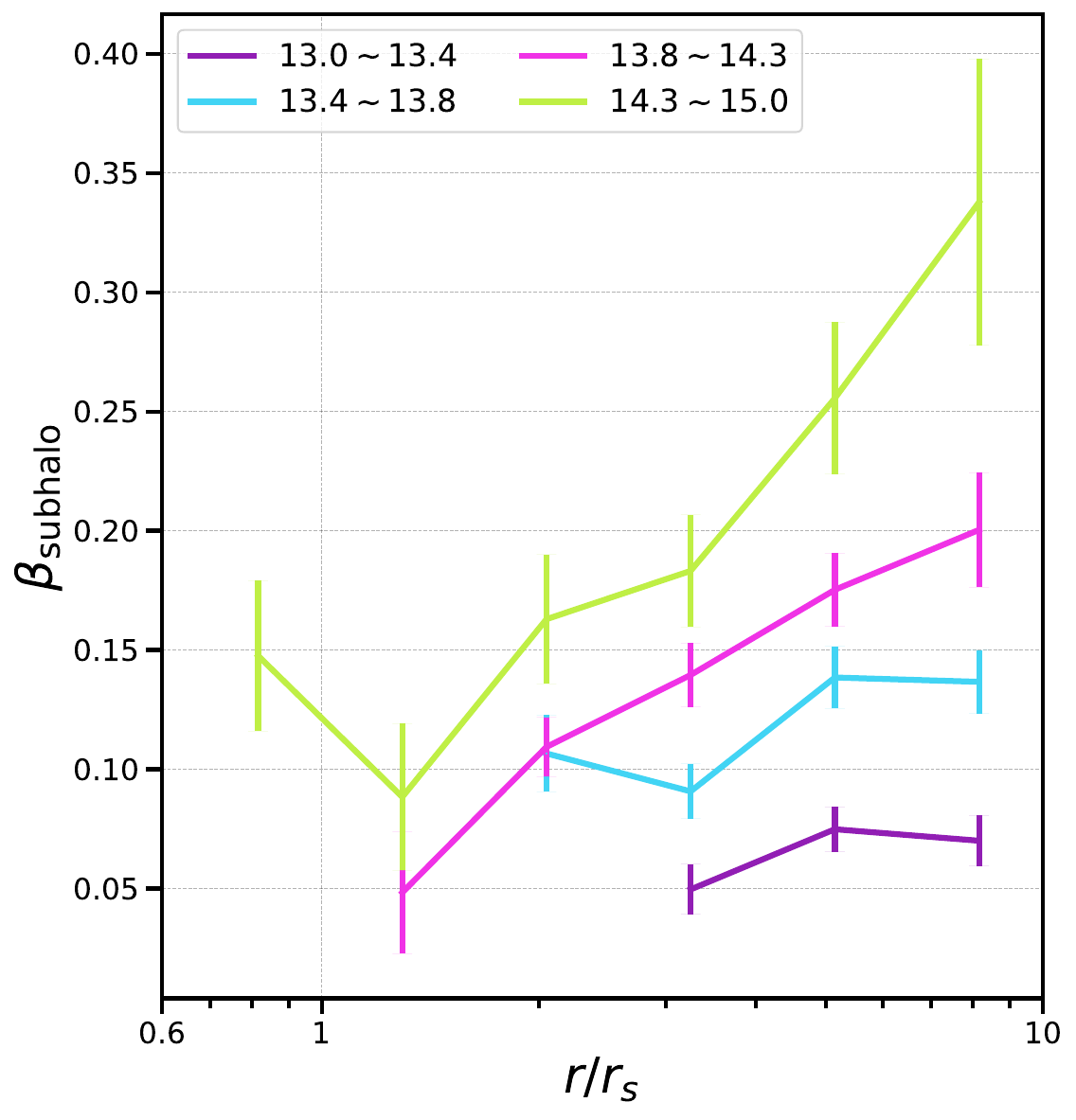}
\caption{Similar to Figures~\ref{fig:beta_star_rs} and \ref{fig:beta_dm_rs} above, but for subhalos from TNG300. Wide radial bins are adopted for subhalos, due to their much smaller sample size.}
\label{fig:beta_sub_rs}
\end{figure}

\section{results}
\label{sec:results}

\subsection{Velocity anisotropy profiles of halo stars, dark matter and subhalos: dependence on halo mass}
\label{sec:baryon}

Figure~\ref{fig:beta_star_rs} shows the velocity anisotropy profiles for halo stars in galaxy systems binned according to their host dark halo virial mass, $M_{200}$. $r$ in the $x$-axis is the galactocentric radius. We scale it by the scale radius, $r_s$, of the Navarro-Frenk-White \citep[][NFW]{1996ApJ...462..563N,1997ApJ...490..493N} profile. Explicitly, $r_s$ for each system is obtained by fitting the NFW model profile to the density profile calculated from all particles in the simulation. 

Figure~\ref{fig:beta_star_rs} shows that $\beta$ generally becomes more radial with the increase in the galactocentric radius, $r$. For the three more massive bins, the curves reach maximums at $r\sim 3-5~r_s$, beyond which the curves start to drop slightly\footnote{The number density profiles of the stellar halo drops very fast, and thus results beyond $r\sim 3-5~r_s$ would suffer from larger statistical uncertainties due  to the small number of halo stars at large radii, so we avoid over interpreting the decrease in $\beta$ at large radii.}. At outer radii of $r>\sim r_s$, we can see a prominent trend that $\beta$ becomes more radial with the increase of $M_{200}$. This is true except for the green curve, which starts to drop beyond 3~$r_s$, but this might be due to the limited number of massive galaxy systems in this bin and the small number of halo stars beyond 3~$r_s$, and thus the measurements suffer from large uncertainties. On the other hand and in more inner regions ($r<\sim r_s$), $\beta$ instead decreases with the increase of $M_{200}$. 

Before discussing what causes the change of the velocity anisotropy of halo stars with the increase of $M_{200}$, and why it behaves differently in inner and outer radii, we first show similar versions of Figure~\ref{fig:beta_star_rs} for dark matter particles from TNG300 (Figure~\ref{fig:beta_dm_rs}), and for subhalos (Figure~\ref{fig:beta_sub_rs}). Note the dark matter particles are from TNG300, and we postpone discussions on the comparison between TNG300 and TNG300-Dark to the next subsection (Section~\ref{sec:baryon}).

For dark matter particles, their $\beta$ gets more radial with the increase in $r$, consistent with previous studies \citep[e.g.][Meng et al. in preparation]{2005MNRAS.361L...1W,2008MNRAS.386.2022A,2012ApJ...752..141L}. $\beta$ for dark matter particles also monotonically gets more radial with the increase of $M_{200}$ at $r>\sim0.3~r_s$, and the trend is more prominent at larger $r$. This trend will also be presented in Meng et al. in preparation. The trend is similar to the what we see at $r>\sim r_s$ for halo stars in Figure~\ref{fig:beta_star_rs}, but it is much weaker for halo stars, whereas it is more prominent for dark matter particles in Figure~\ref{fig:beta_dm_rs}. Within $\sim0.3~r_s$ of Figure~\ref{fig:beta_dm_rs}, the trend for dark matter particles is similar to what we see at $r<\sim r_s$ for halo stars in Figure~\ref{fig:beta_star_rs}, i.e., $\beta$ becomes more isotropic with the increase of $M_{200}$. 

Subhalos demonstrate a similarly prominent trend at outer radius in Figure~\ref{fig:beta_sub_rs}, that $\beta$ for subhalos gets more radial with the increase of $M_{200}$ as well. This trend also gets more prominent at larger $r$. However, because the number of subhalos is very low in central regions due to tidal disruptions, the measurements in more central regions of less massive bins do not satisfy our requirement that more than 80\% of galaxy systems can have more than 5 subhalos in a given radial bin (see Section~\ref{sec:method}), we cannot push robust measurements to more inner regions within $r_s$ for the cyan and purple curves. We cannot have robust measurement for the least massive bin of $12.6<\log_{10}M_{200}/\msun<13.0$ as in previous plots either, so this bin is not shown. 

\begin{figure}
\centering
\includegraphics[width=1\linewidth]{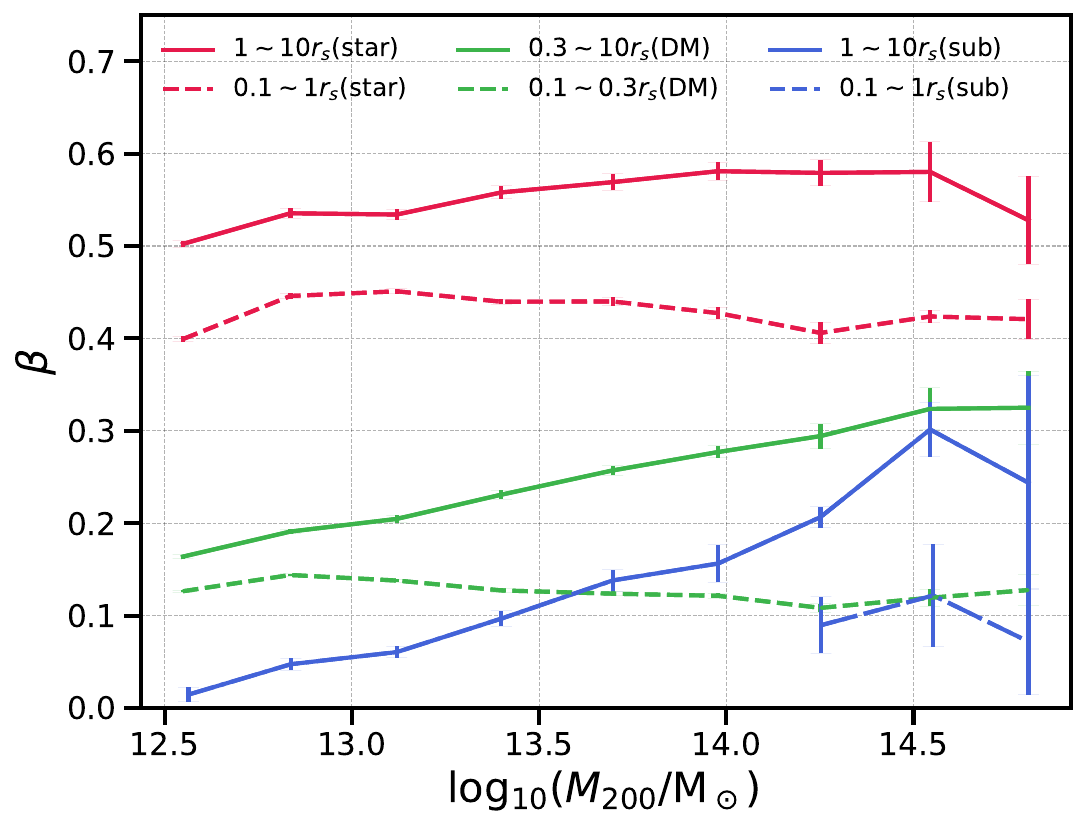}
\caption{Relations between the velocity anisotropy parameter and the average virial mass of host dark matter halos ($M_{200}$). Red, Green and blue curves are based on halo stars, dark matter particles and subhalos, respectively. Solid curves are based on objects at outer radii, while dashed curves are based on objects in more inner regions (see the legend). The errorbars are 1-$\sigma$ scatters of 100 bootstrapped subsamples of galaxies in the same bin of $M_{200}$.}
\label{fig:beta_glo}
\end{figure}

Combining the results for halo stars, dark matter particles and subhalos, we can conclude that they share common features, whereas also have delicate and prominent differences. To more clearly compare the three different populations of objects, we summarize the results in Figure~\ref{fig:beta_glo}, where we show the velocity anisotropies calculated using halo stars, dark matter particles and subhalos in two different radial ranges. Dashed lines refer to the velocity anisotropies calculated using particles in more inner regions, while solid lines are based on those objects at outer radii. The inner and outer divisions are chosen as $r_s$, $0.3~r_s$ and $r_s$ for halo stars, dark matter particles and subhalos, determined empirically according to the radii where the trends change in Figures~\ref{fig:beta_star_rs}, \ref{fig:beta_dm_rs} and \ref{fig:beta_sub_rs}. Similar to how we calculate the profiles in previous figures, we report the median velocity anisotropy ($y$-axis) as a function of the average $M_{200}$ ($x$-axis). Note here the divisions for inner and outer regions are different between halo stars and dark matter particles. If we choose the division of $r\sim r_s$ for dark matter particles, the inner trend can be affected. We adopt different divisions here, but we will show later in Figure~\ref{fig:beta_1-10rs} of how $\beta$ changes with $M_{200}$ in the outer region based on the same radial range, for halo stars, dark matter particles and subhalos. However, the fine tuning in the chosen radial range barely affects the trends in the outer radial region.

According to Figure~\ref{fig:beta_glo}, it is very clear that all solid curves show monotonic increase in $\beta$ with the increase of $M_{200}$. The trend is the most prominent for subhalos, which becomes weaker for dark matter particles, and is the least prominent for halo stars. Here we can go to a smaller mass range for the blue solid curve of subhalos, compared with Figure~\ref{fig:beta_sub_rs} above, simply because the broad radial range includes larger number of subhalos. On the contrary and in inner regions (dashed curves), halo stars and dark matter particles share similar trends that $\beta$ decreases with the increase in $M_{200}$. In inner regions, because the number of subhalos is too small, lower $M_{200}$ bins do not satisfy the criterion of the lower particle number limit in Section~\ref{sec:method}, so we only have measurements of three most massive data points, which barely lead to any prominent trend. 

The amplitudes of $\beta$ based on different types of objects in Figure~\ref{fig:beta_glo} also differ. Halo stars have the largest $\beta$ (the most radial) at all $M_{200}$, and the velocity anisotropy of subhalos are in general more isotropic than those of halo stars and dark matter particles. The difference between dark matter particles and subhalos is generally consistent with \cite{2004MNRAS.352..535D}. Moreover, according to all previous figures, it is clear that the outer halo region is more radial than the inner region, for all types of objects/particles probed so far. 

We now try to discuss why different types of objects have different amplitudes in $\beta$, and discuss possible reasons causing the different behaviors of how $\beta$ changes with the increase of $M_{200}$. We first discuss the amplitude difference. As we all know, the stellar and dark matter halos both grow through accretion. Halo stars and dark matter particles are stripped from infalling satellite galaxies or subhalos. This would introduce a selection bias, that those stripped halo stars and dark matter particles\footnote{Note when calculating the velocity anisotropy for ex-situ halo stars and dark matter particles, we have excluded those particles bound to surviving subhalos/satellites.} come more from those subhalos/satellites that have more radial orbits, hence passing close or even have gone through the center of the host system, whose stellar and dark matter particles can then be more quickly and easily stripped. This explains why halo stars and dark matter particles have more radial $\beta$ than that of subhalos, because they more likely come from those subhalos having more radial orbits. Besides, before getting stripped, halo stars are the most bound part in the very center of subhalos hosting these satellite galaxies. They are more difficult to be stripped than those less bound dark matter particles in outskirts of subhalos. This introduces stronger selection bias for halo stars, i.e., those halo stars come from satellites/subhalos with even more radial orbits than those of stripped dark matter particles.

A similar trend has been found by \cite{2015MNRAS.453..377W} (see their Figure~5). Note the sample of halo stars used in \cite{2015MNRAS.453..377W} was not based on hydrodynamical simulations. Instead, the halo stars are those most bound dark matter particles in subhalos that are tagged to represent halo stars \citep{2010MNRAS.406..744C}. However, the physics behind is the same. Halo stars were the most bound part of subhalos before getting stripped, and they more likely come from satellites/subhalos with more radial orbits, hence showing the most radial $\beta$. 

\begin{figure}
\centering
\includegraphics[width=1\linewidth]{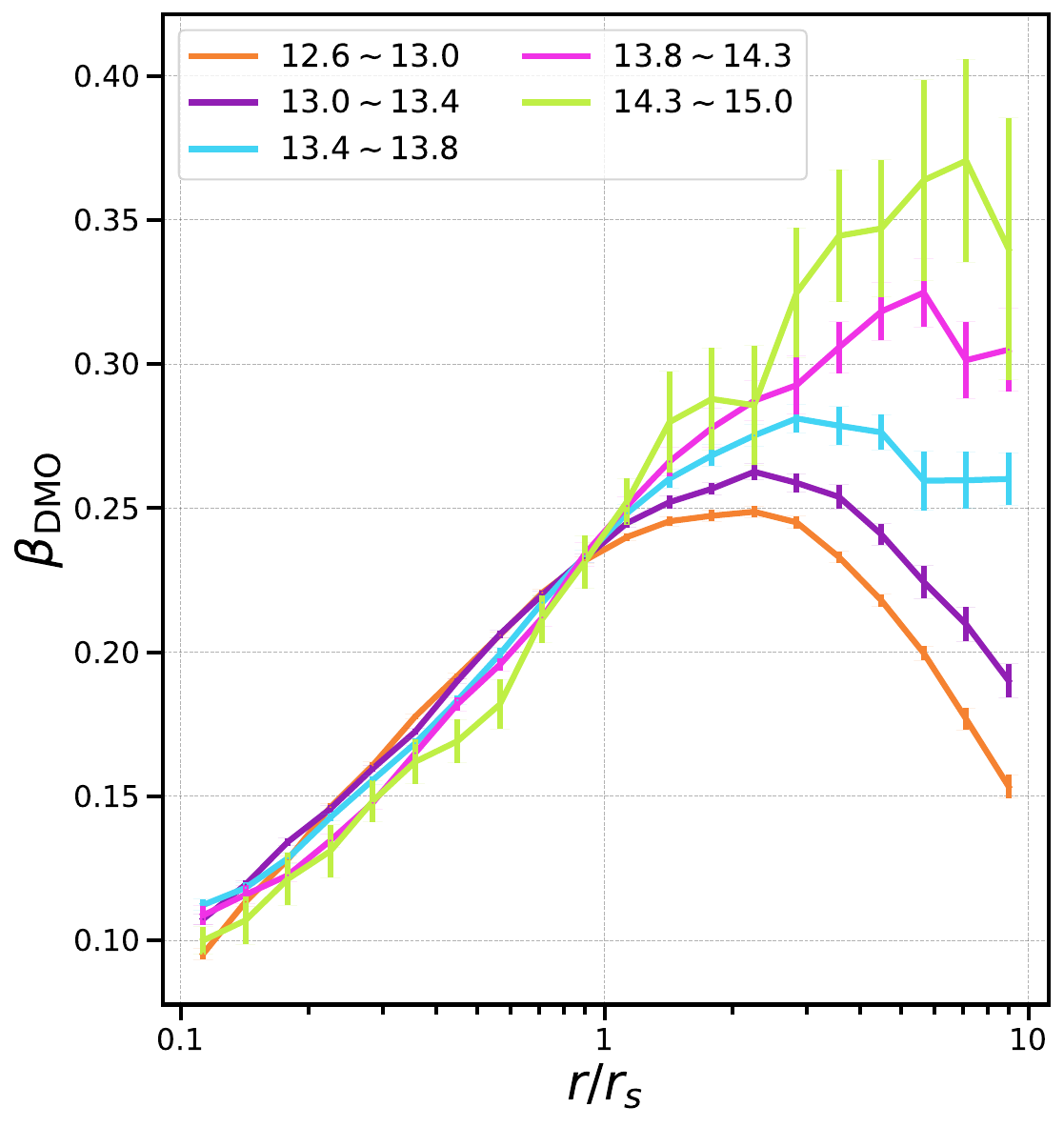}
\caption{Similar to Figure~\ref{fig:beta_dm_rs}, but for dark matter particles from TNG300-Dark. }
\label{fig:beta_dmo_rs}
\end{figure}

\begin{figure*}
\centering
\includegraphics[width=0.9\linewidth]{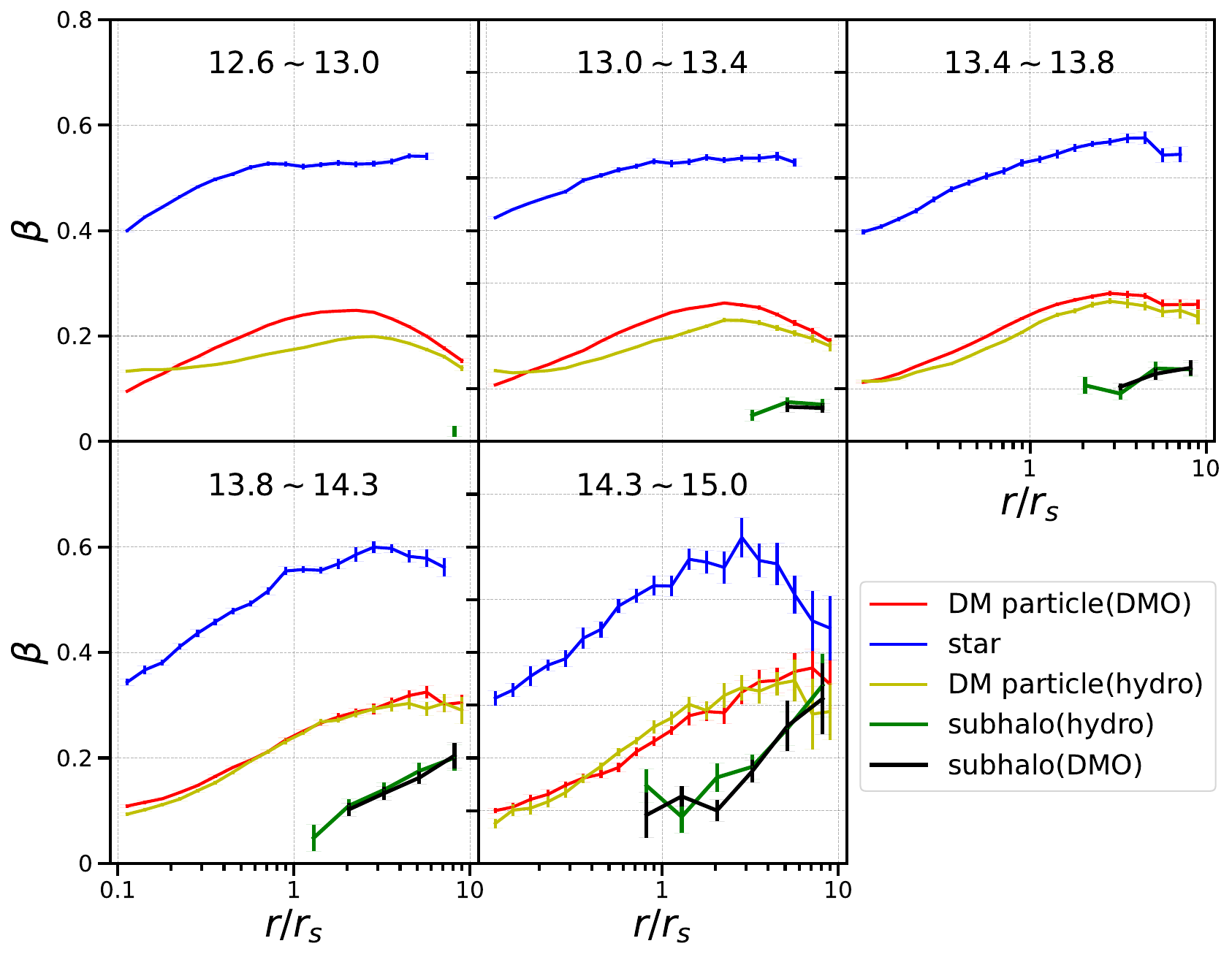}
\caption{The velocity anisotropy profiles for halo stars (blue), dark matter particles from TNG300 (yellow) and TNG300-Dark (red) and subhalos from TNG300 (green) and TNG300-Dark (black). Each panel refers to a given mass bin in $\log_{10}M_{200}/\msun$ (see the text in each panel). The number of subhalos is too few to have good measurements at $r<r_s$, and this is the reason for the truncation of the green and black curves. }
\label{fig:beta_bins}
\end{figure*}

\begin{figure}
\centering
\includegraphics[width=1\linewidth]{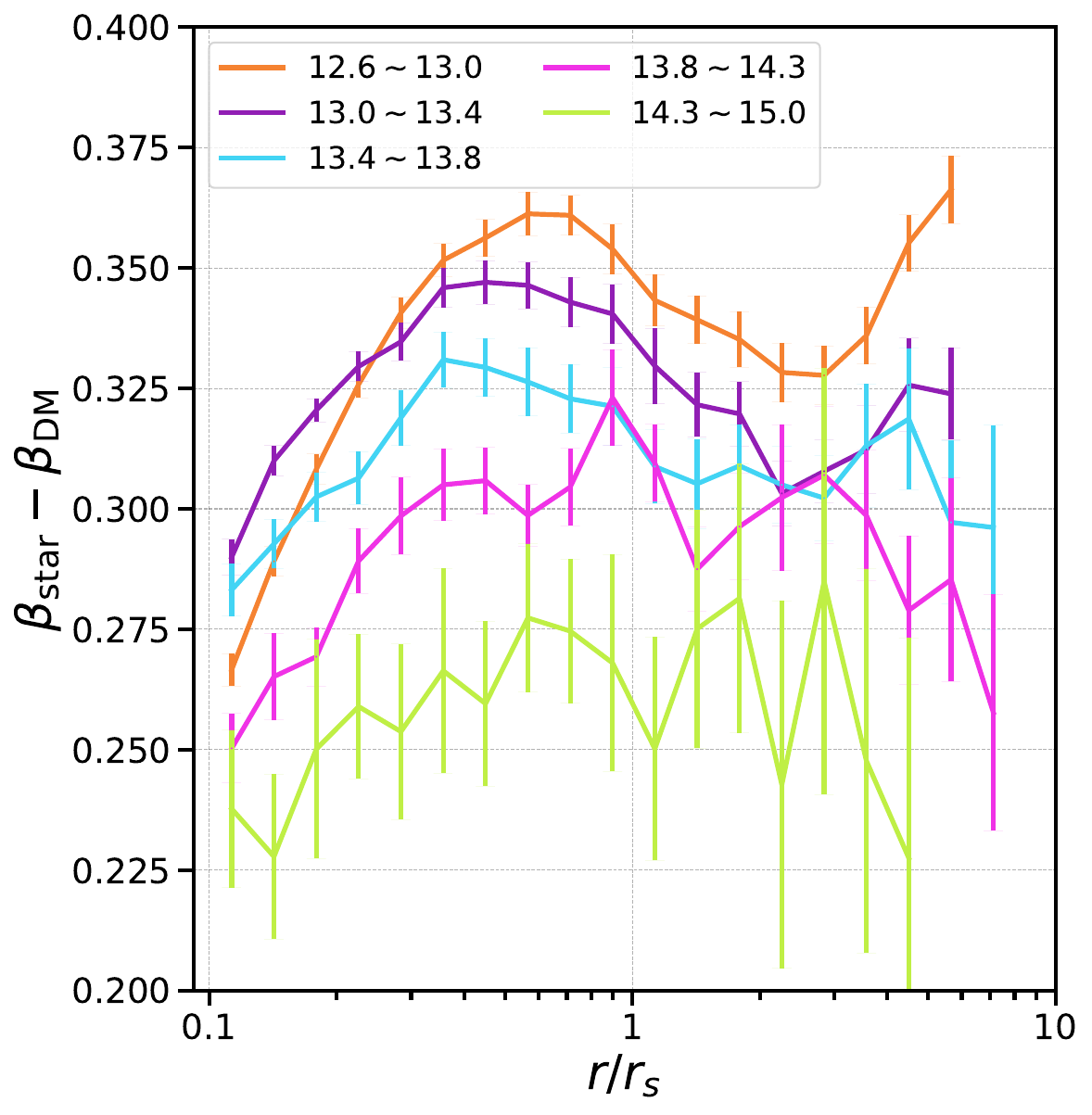}
\caption{The difference between the velocity anisotropy profiles of halo stars and dark matter particles, for different bins of $M_{200}$ (see the legend for the range in $\log_{10}M_{200}/\msun$).}
\label{fig:beta_star-dm}
\end{figure}

We now discuss possible physical causes for the change in $\beta$ with $M_{200}$. First of all, we have seen that halo stars, dark matter particles and subhalos all show monotonic increase of their $\beta$ with the increase in $M_{200}$ at outer radii. We think this trend is related to the infalling orbits of accreted satellites/subhalos. In a previous study, \cite{2020ApJ...905..177L} investigated the infalling orbital distribution of satellite galaxies or subhalos using a large sample of cosmological simulations. They found that the magnitude of infalling velocities for satellites/subhalos follows a nearly universal distribution, but subhalos that are hosted by more massive halos tend to move along more radial directions. This would explain the monotonic dependence of $\beta$ on $M_{200}$ at outer radii shown from Figure~\ref{fig:beta_star_rs} to Figure~\ref{fig:beta_glo}. If satellites/subhalos around more massive halos fell in along more radial orbits, we expect them to have larger radial velocities than tangential velocities. Combining satellites/subhalos fell in at different times, this would introduce larger radial velocity dispersions, hence larger $\beta$. \cite{2020ApJ...905..177L} further reported that their findings are nearly independent of redshifts when the density peak height is used as the proxy for host halo mass, and the results are consistent with the scenario where the dynamical environment is relatively colder for more massive structures because their own gravity is more likely to dominate the local density field. 

Intuitively, it is also easy to understand why high mass halos have more radial orbits in the outer part, by considering the evolution phase of halos. In the hierarchical universe, large halos form later and lagging behind the evolution of low mass halos. \citet{FH21} has shown that galactic size halos have largely terminated their mass accretion with no prominent infall region around them, i.e., effectively \emph{depleted} their infall region, while high mass halos are surrounded by an active infall region with prominent radial motions~\citep[see also][]{Gao23,Diemer22}. Thus it is natural to expect that the orbits in the outer halo become more radial in more massive halos. This trend mainly happens at larger radii, because objects in outer halo regions maintain better their infalling orbits, and the orbits become more circular and virialized due to the splashback process~\citep{FG84} which gradually decreases the pericenter distances of the orbits with time passes.

%With the findings and the scenario proposed by \cite{2020ApJ...905..177L}, it is natural to expect the monotonic increase of $\beta$ with the increase of $M_{200}$. 
The trend is the weakest for halo stars, and the strongest for subhalos, with that of dark matter particles in between. This is also natural to understand, because as we have mentioned above, halo stars and dark matter particles are stripped from satellites and subhalos, which must have fell in at an earlier time in order for mass stripping to operate. The stripped material thus have become more circularized due to the splashback process mentioned above. %The stripped part come from a more radially biased subsample. 
 This may weaken the trend with respect to all parent satellites/subhalos. Moreover, simply because the allowed range of change in $\beta$ is limited to higher values closer to unity, the narrower range of change makes the trend weaker.

In more inner regions, satellites/subhalos have lost memories about the initial infalling orbits. Moreover, satellites/subhalos with more radial orbits might be depleted due to the strong tidal effects in the center. Thus we see different trends in inner and outer radii. In more inner regions, the trend is similar between halo stars and dark matter particles, that $\beta$ decreases in more massive systems, though for dark matter particles this happens in more inner radii of $r<0.3~r_s$ and this is not clearly seen for subhalos. 

The inner trend is very likely affected by the in-situ component. Usually, more massive galaxies are dominated by elliptical/spheroidal morphology, whose member stars are dominated by random motions and are thus expected to have more isotropic orbital distributions \citep[e.g.][]{1976ApJ...204..668F,2002MNRAS.332..901N,2014MNRAS.442.3284A}. Though in our calculations, only ex-situ star particles are used, those ex-situ stars in the very inner region may have exchanged energy and angular momentum through dynamical interaction with the in-situ particles, hence also showing more isotropic $\beta$. Besides, since there are more dark matter particles than star particles, the inner trend for dark matter particles may start to happen at even smaller radii, where the in-situ component can be more dominant over dark matter.

In the next subsection, we further investigate the effect of baryons, by comparing the velocity of dark matter particles in TNG300 and TNG300-Dark. If our argument about the inner trend is correct, we would expect very different behavior in the inner trend for dark matter particles from TNG300-Dark, compared with TNG300.

\subsection{Effect of baryons on velocity anisotropy}
\label{sec:baryon}

Figure~\ref{fig:beta_dmo_rs} shows a similar version of the velocity anisotropy profiles for dark matter particles, as in Figure~\ref{fig:beta_dm_rs}, but this is based on TNG300-Dark. Comparing Figures~\ref{fig:beta_dmo_rs} and \ref{fig:beta_dm_rs}, it is very clear that in TNG300-Dark, $\beta$ profiles of dark matter particles show very similar dependence on $M_{200}$ beyond $r\sim r_s$ as in the hydro version. However, at $r<\sim r_s$, the trend is significantly different. In the DMO simulation, different color curves almost overlap with each other within $r=r_s$, showing at most some quite weak decrease in $\beta$ with the increase of $M_{200}$. A similar result has been found and will be available in Meng et al. (in preparation). This supports our argument above that the inner trends in Figures~\ref{fig:beta_star_rs} and \ref{fig:beta_dm_rs} are related to the in-situ baryons. 

We also note that while we see monotonic increase in $\beta$ with the increase in $M_{200}$ for dark matter particles from TNG300 at $0.3~r_s<r<10~r_s$ (Figure~\ref{fig:beta_dm_rs}), which we attribute to the more prominent infalling motion of satellites/subhalos around more massive dark matter halos \citep{2020ApJ...905..177L,FH21}, this trend is absent for dark matter particles from TNG300-Dark in Figure~\ref{fig:beta_dmo_rs} over $0.3~r_s<r<r_s$. One possibility causing this is that subhalos in the hydro TNG300 simulation is more self-bound and resistant to tidal effects due to the existence of baryons~\citep[e.g.,][]{Sifon}. In other words, the presence of baryons helps to delay the stripping of subhalos until they get much closer to the center of the halo than in the DMO run, hence modifying the behavior of $\beta$ over $0.3~r_s<r<r_s$. We will return to discuss this point later in this subsection.

This is more clearly shown in Figure~\ref{fig:beta_bins}, where we plot the velocity anisotropy profiles for halo stars,  dark matter particles and subhalos from both TNG300 and TNG300-Dark together. Each panel refers to a given mass range of $\log_{10}M_{200}/\msun$, as indicated by the text in the panel. The difference between dark matter particles from TNG300 and TNG300-Dark not only happens at $r<r_s$, as we have already seen in the previous Figures~\ref{fig:beta_dm_rs} and \ref{fig:beta_dmo_rs}, but also exists at larger radii, that dart matter particles from the hydro simulation tend to be slightly more isotropic than those from the DMO version over the radial range of $0.2~r_s<r<10~r_s$. In particular, this difference mainly exists for smaller systems with $\log_{10}M_\ast/\msun<\sim13.8$. Our findings are consistent with a previous study based on the Illustris simulation \citep{2019MNRAS.484..476C}, in which dark matter particles in hydro version of simulations are also found to be more isotropic than those in the DMO version. In the very center for the two least massive bins of $\log_{10}M_\ast/\msun<\sim13.4$, $\beta$ becomes more radial in the hydro simulation than the DMO simulation.

The fact that dark matter particles in the hydro run is slightly more isotropic at $0.2~r_s<r<10~r_s$ and $\log_{10}M_\ast/\msun<\sim13.8$ may be related to the more isotropic potential, as affected by the existence of more homogeneously distributed hot gas, for example, hence resulting in more isotropic orbital distributions. Another possibility, as we have already mentioned above, is that the existence of baryons would make the infalling subhalos more resistant to tidal effects at larger radii, thus reducing the radial population in the stripped particles. In other words, particles which would have already been stripped from subhalos with radial orbits at larger radii in the DMO version, may still maintain bound to their parent subhalos in the hydrodynamical version and then eventually get stripped in more central regions. Also we would expect more subhalos with radial orbits still stay bound and survive. 
However, these subhalos would eventually get stripped as they approach the denser and more central regions of the halo, and thus increase the radial population in the stripped particles at the very center. This can result in the flattening of the $\beta$ profiles of dark matter particles in hydro simulations as we have seen in Figure~\ref{fig:beta_bins}. Comparing the green and black curves in Figure~\ref{fig:beta_bins}, we can see the velocity anisotropy of bound/survived subhalos is slightly more radial in the hydro version, but the errorbars of the green and black curves are much larger than those of the other curves, preventing us from seeing significant differences. 

We note for the two most massive panels of Figure~\ref{fig:beta_bins}, the trend is slightly opposite compared with other less massive panels. In the very center, $\beta$ is slightly more isotropic in the hydro run than in the DMO run at $\log_{10}M_{200}/\msun>13.8$, and is slightly more radial at $0.3~r_s<r<2.5~r_s$ and $\log_{10}M_{200}/\msun>14.3$. As we have mentioned with Figures~\ref{fig:beta_dm_rs} and \ref{fig:beta_dmo_rs} above, massive galaxies are dominated by elliptical/spheroidal morphology with isotropic stellar orbits, and the dark matter particles in the very inner region may have exchanged energy and angular momentum through dynamical interaction with the in-situ component, hence becoming more isotropic.

In all panels of Figure~\ref{fig:beta_bins}, we can clearly see that the $\beta$ profile for halo stars is the highest at all radii, while the $\beta$ for subhalos is the lowest, with the curves for dark matter particles in between. Blue curves for halo stars are approximately shifted up by a constant, which shows a weak dependence on radius. We thus further show in Figure~\ref{fig:beta_star-dm} the difference between the velocity anisotropy of stars and dark matter particles, as a function of radius and for different $M_{200}$. For a given bin of $M_{200}$, the difference is smaller at smaller galactocentric radius, which gradually increase with the increase of $r$, reaching a maximum at $r\sim 0.5-1~r_s$, and then start to be closer to a constant at $r>r_s$. From the highest to lowest $M_{200}$ bins, the values go from $\sim$0.27 to $\sim$0.35 at $r>r_s$. This is an interesting point, suggesting that if we can precisely measure the $\beta$ profile for halo stars beyond $r_s$, we can use our conclusion to infer the $\beta$ profiles for dark matter beyond $r_s$ by roughly adding this value based on the simulation prediction. In Section~\ref{sec:ext} and the Appendix below, where we extend our measurements down to smaller halo masses, we provide a fitting formula describing the difference between the $\beta$ for halo stars and for dark matter particles, for practical purpose.

\begin{figure}
\centering
\includegraphics[width=1\linewidth]{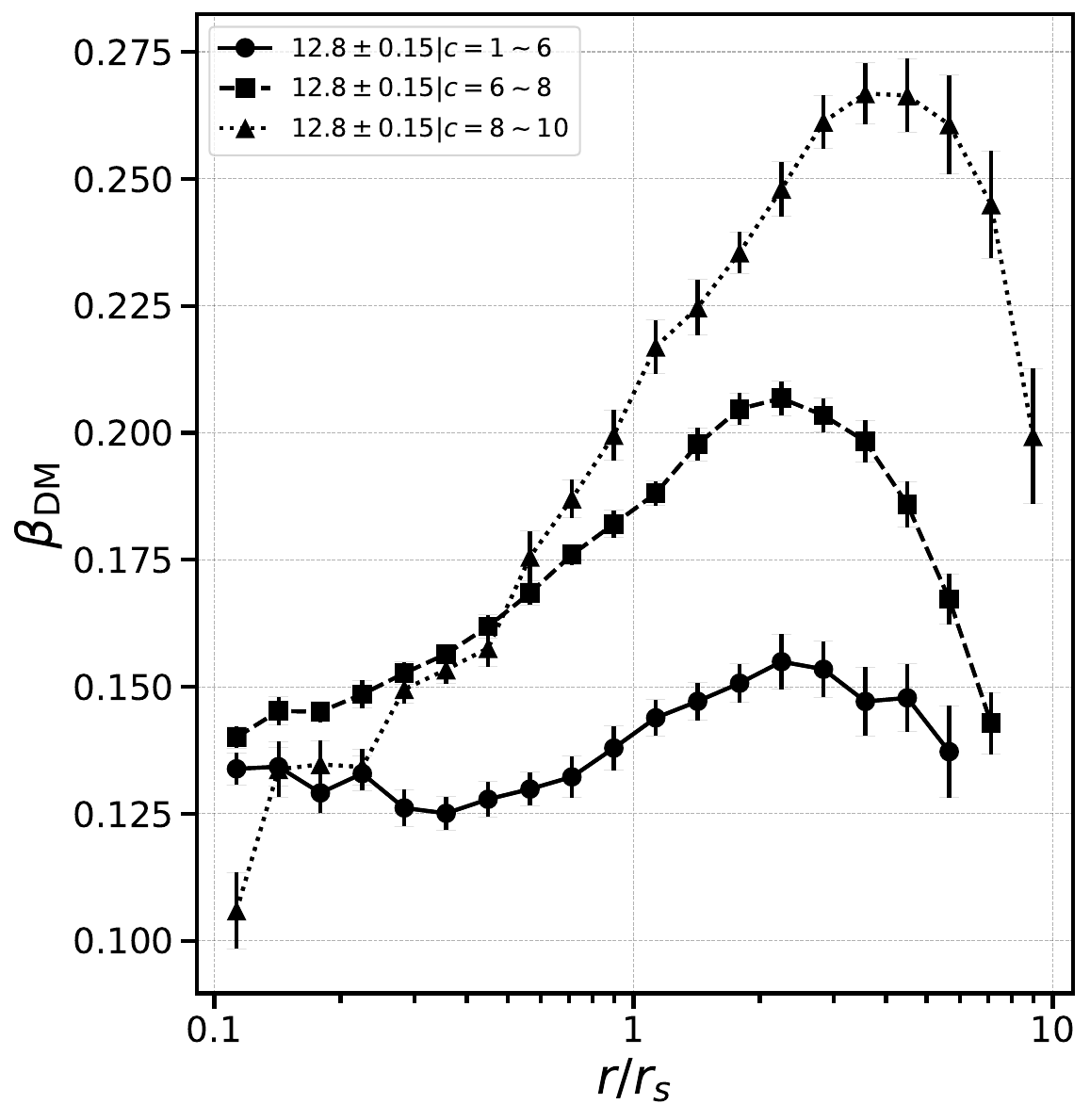}
\caption{Velocity anisotropy profiles of dark matter particles from TNG300. Curves with the same color are based on systems in the same mass range of $\log_{10}M_{200}/\msun$, as indicated by the first few numbers in the legend. We further divide them into three different bins of halo concentration ($c_{200}$), also indicated by the legend. The errorbars are 1-$\sigma$ scatters of 100 bootstrapped subsamples of galaxy systems in the same mass bin.}
\label{fig:beta_dm_mc}
\end{figure}

\begin{figure}
\centering
\includegraphics[width=1\linewidth]{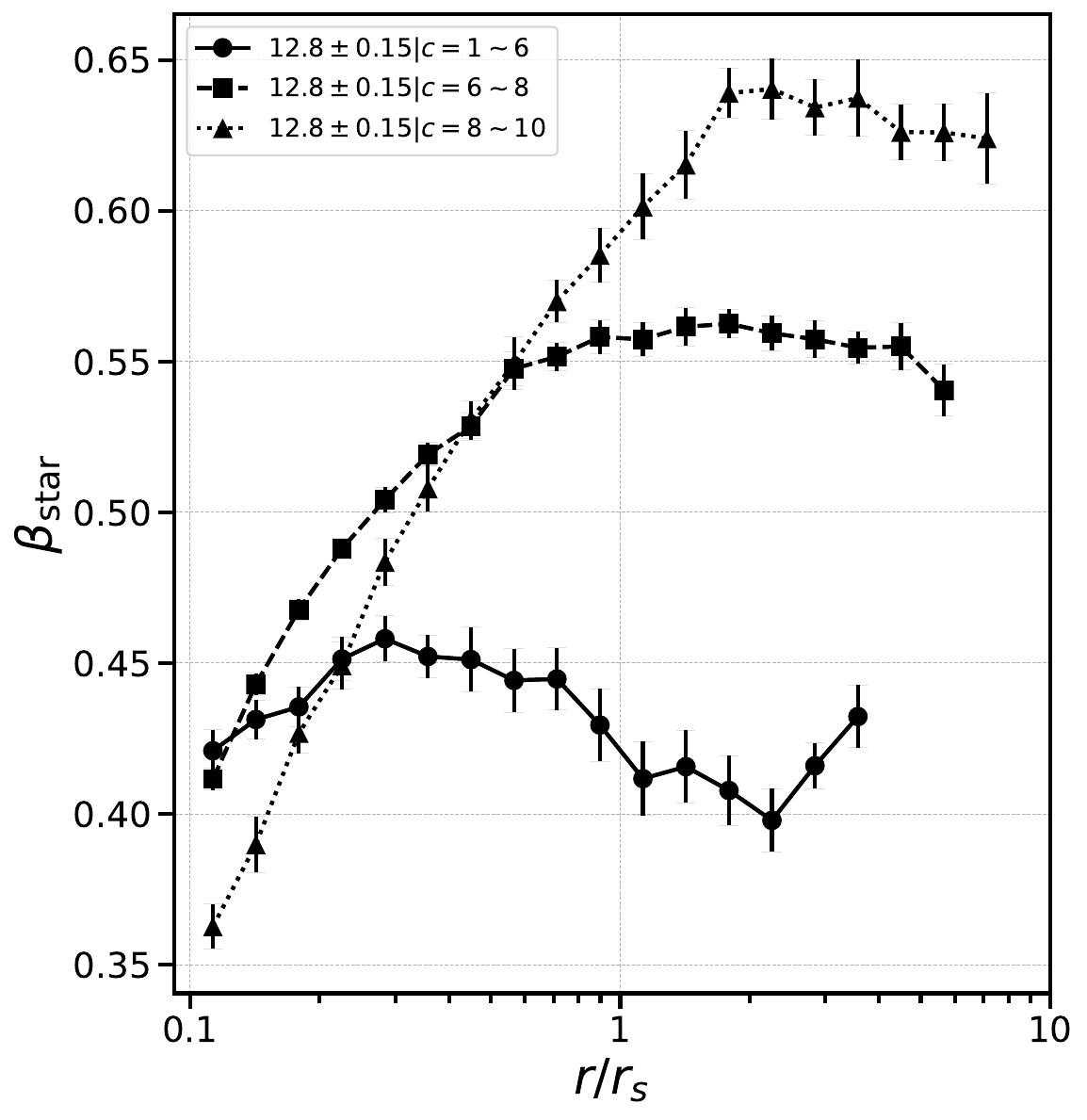}
\caption{Similar to Figure~\ref{fig:beta_dm_mc}, but is based on halo stars.}
\label{fig:beta_star_mc}
\end{figure}

\begin{figure*}
\centering
\includegraphics[width=0.49\linewidth]{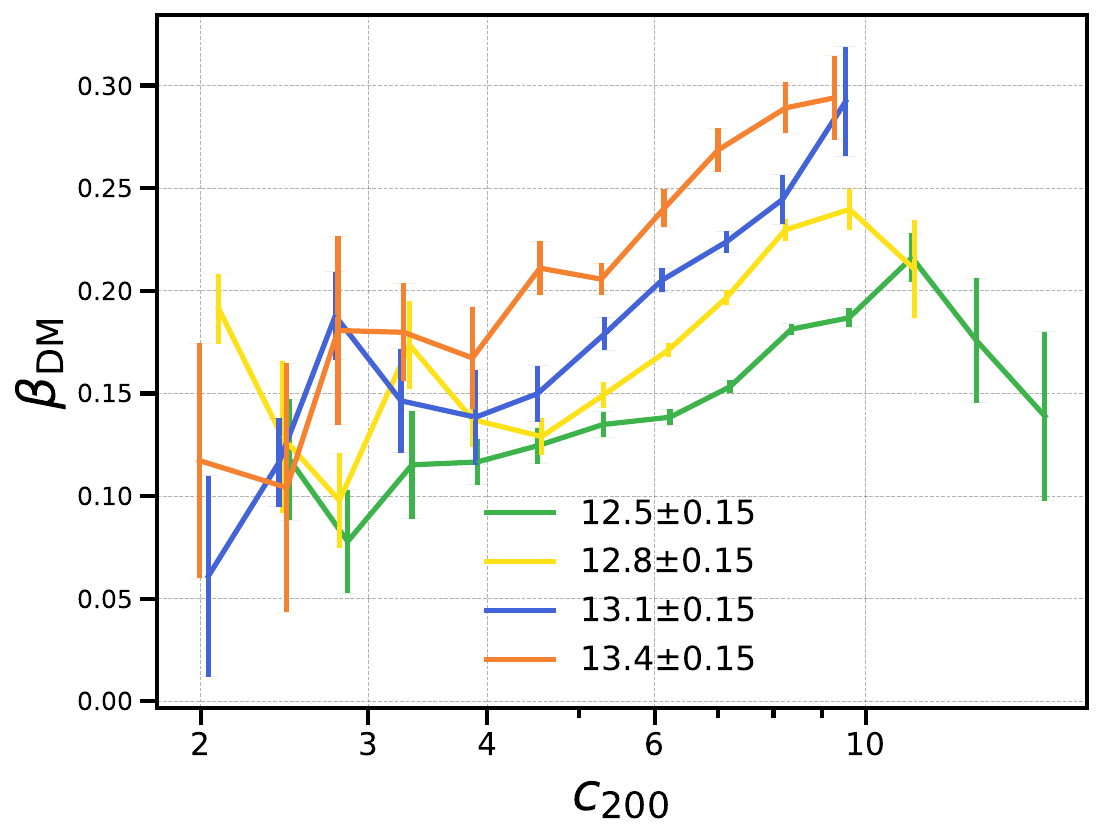}
\includegraphics[width=0.49\linewidth]{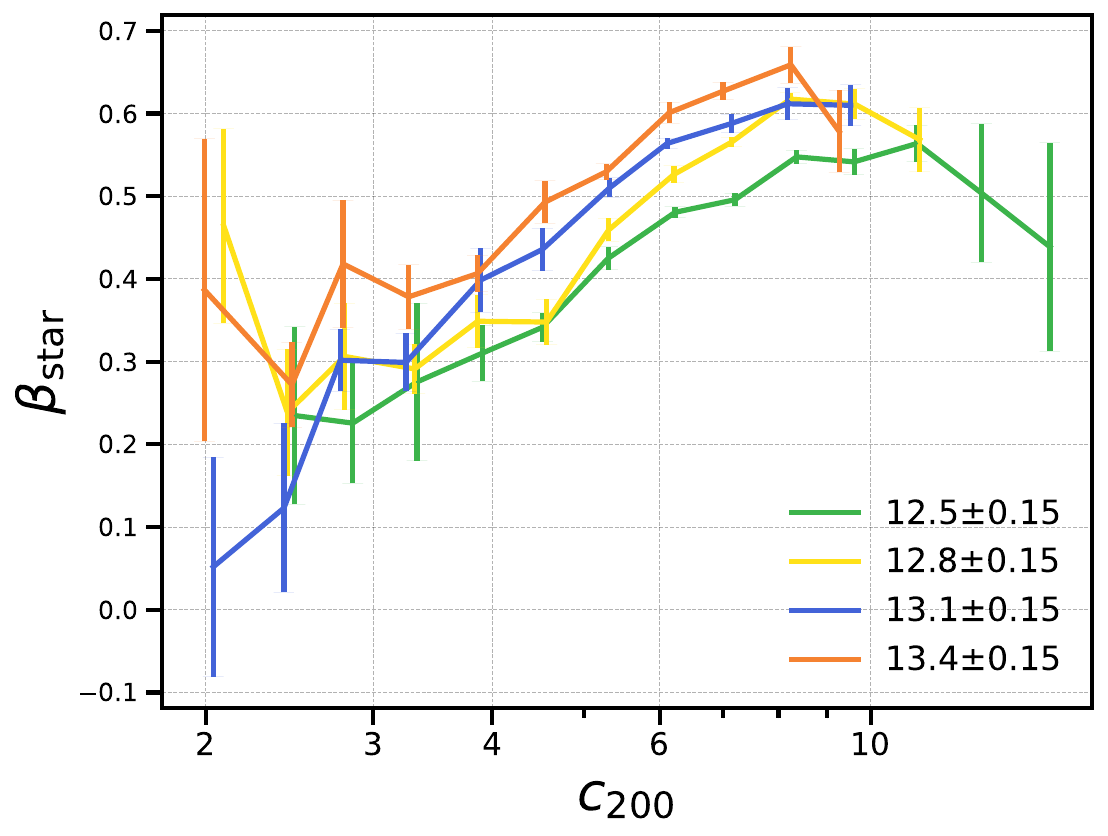}
\caption{The velocity anisotropy ($\beta$) as a function of averaged concentration parameter ($c_{200}$) of host dark matter halos, for galaxies in the same bin of host halo virial mass, $M_{200}$. Each color curve is based on a given mass bin of $M_{200}$, with the legend showing the ranges in $\log_{10}M_{200}/\msun$. $\beta$ is calculated using dark matter particles (left) with galactocentric radii of $0.3~r_s<r<10~r_s$ or halo stars (right) with $r_s<r<10~r_s$, for galaxy systems from TNG300. Errorbars are 1-$\sigma$ scatters of 100 bootstrapped subsamples of galaxy systems.}
\label{fig:beta_dm_mc_glo}
\end{figure*}

\subsection{Dependence of velocity anisotropy on halo concentration}
\label{sec:concentration}

\begin{figure*}
\centering
\includegraphics[width=0.9\linewidth]{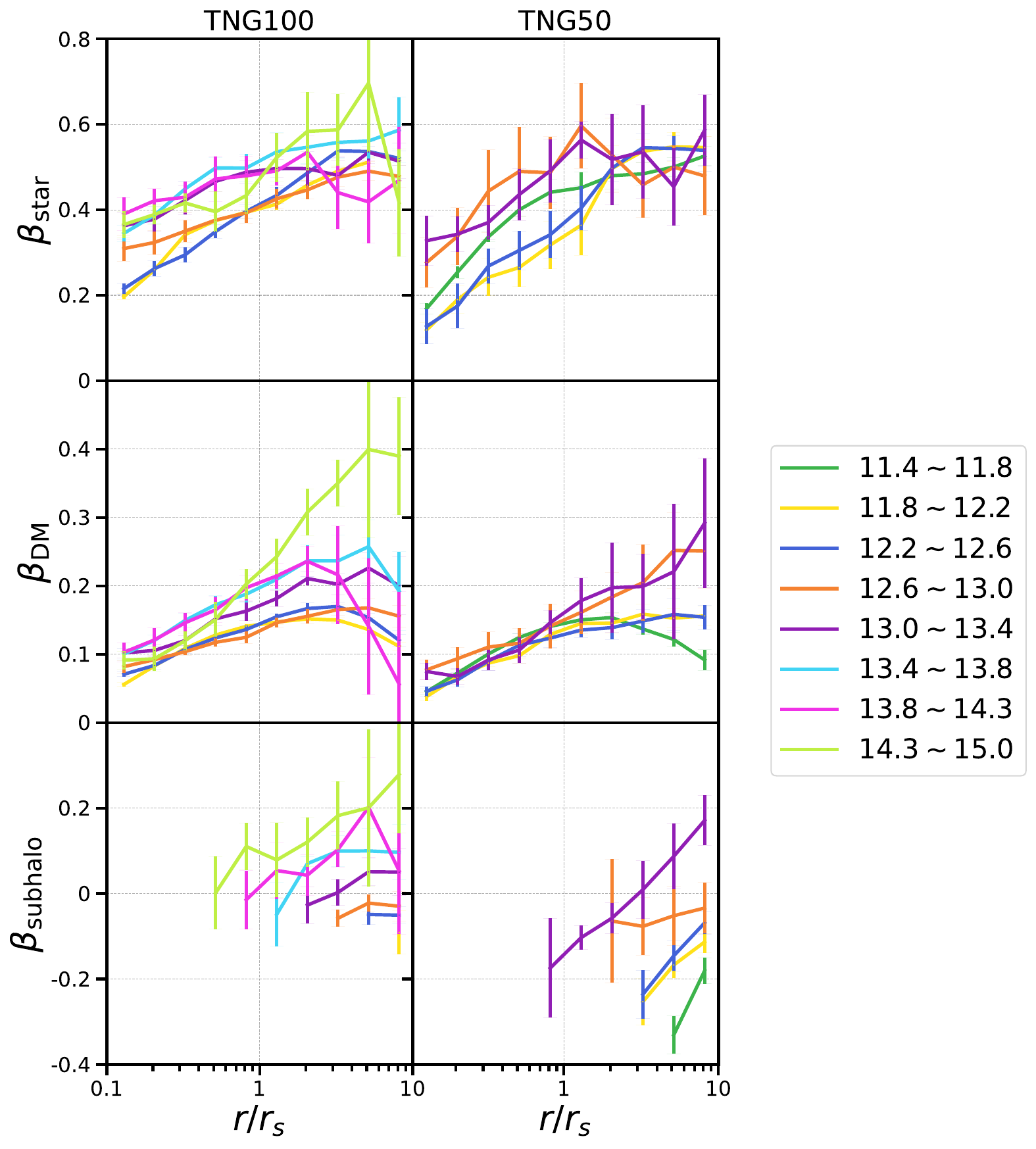}
\caption{Velocity anisotropy profiles for halo stars (top), dark matter particles (middle) and subhalos (bottom) from TNG100 (left) and TNG50 (right). Different color curves refer to different bins in $\log_{10}M_{200}/\msun$, as indicated by the legend. The errorbars are 1-$\sigma$ scatters of 100 bootstrapped subsamples of galaxy systems in the same mass bin. With the higher resolution TNG50 and TNG100 simulations, we can push down to smaller $M_{200}$ than results based on TNG300 in previous figures.}
\label{fig:TNGall}
\end{figure*}

\begin{figure*}
\centering
\includegraphics[width=0.9\linewidth]{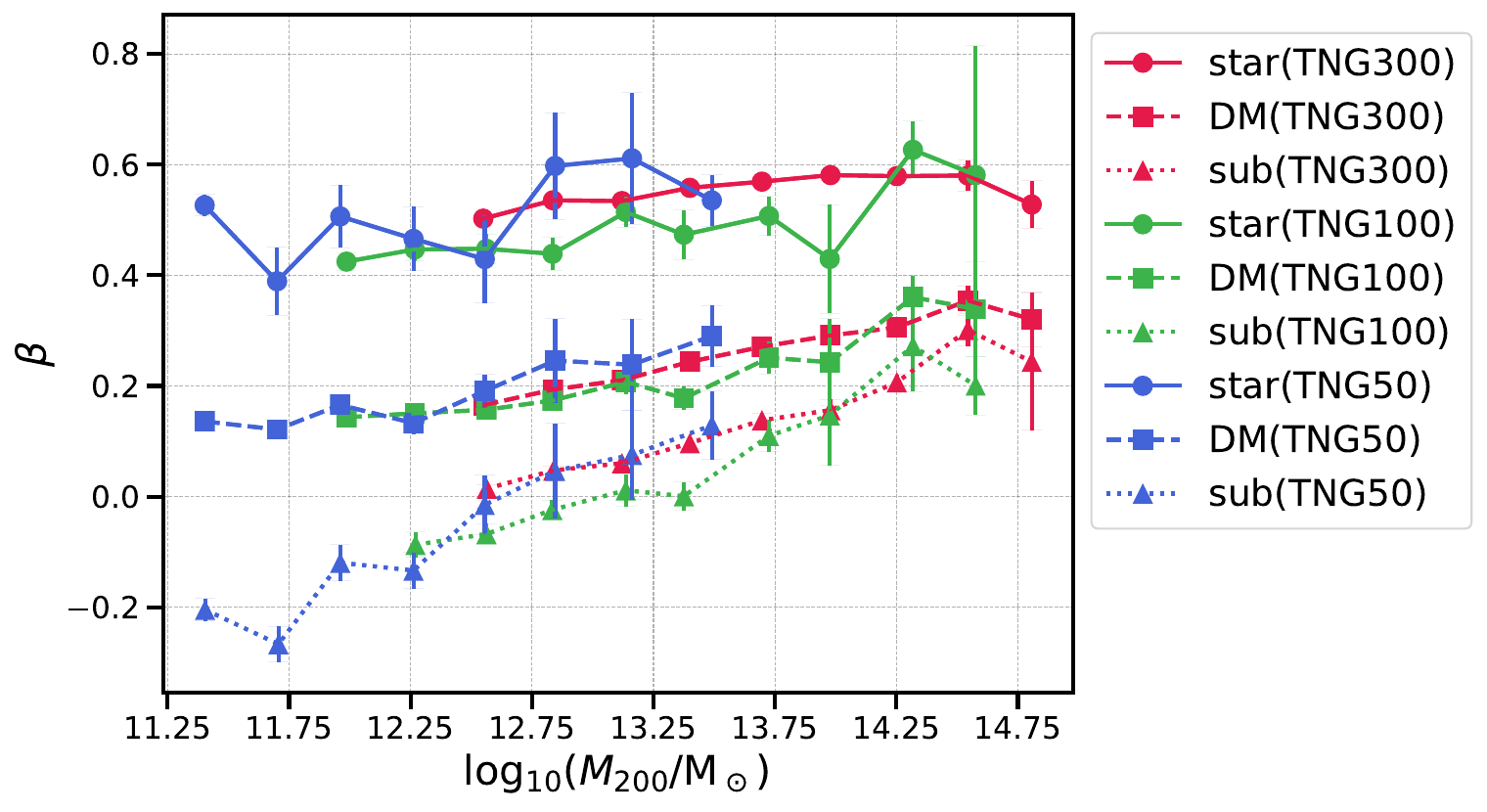}
\caption{Similar to Figure~\ref{fig:beta_glo}, but we have extended the solid curves in Figure~\ref{fig:beta_glo} to smaller mass ranges based on TNG50 and TNG100. Curves in this figure are based on halo stars (solid), dark matter particles (dashed) and subhalos (dotted) with $r_s<r<10~r_s$.}
\label{fig:beta_1-10rs}
\end{figure*}

In previous sections we have investigated how do the velocity anisotropies of halo stars, dark matter particles and subhalos depend on the virial mass of host dark matter halos. Different types of objects show both similar and distinct features in terms of the dependence. In this subsection, we move on to explore the dependence on the other halo parameter, i.e., halo concentration. The halo concentration ($c_{200}$) is defined as the ratio between the virial radius ($R_{200}$) and the scale radius ($r_s$), i.e., $c_{200}=R_{200}/r_s$.

To eliminate the dependence on $M_{200}$, we demonstrate the trend by dividing systems into a narrow halo mass ranges of $12.8\pm 0.15$, and then further divide them into three different bins of $c_{200}=1-6, 6-8$ and $8-10$. The results for dark matter particles from TNG300 are shown in Figure~\ref{fig:beta_dm_mc}. We find that systems having higher concentrations have more radial $\beta$ profiles. The trend is monotonic, with more significant differences at outer radii. Figure~\ref{fig:beta_star_mc} shows a similar version of story for halo stars from TNG300. We see monotonic dependence on $c_{200}$ as well in outer radii.

We show in Figure~\ref{fig:beta_dm_mc_glo} the more clear and general dependence of how $\beta$ changes with $c_{200}$. Different color curves are a few bins of $M_{200}$, and for a fixed bin of $M_{200}$, we further divide galaxies into bins of $c_{200}$. $\beta$ for each system is calculated using dark matter particles (left panel) in the radial range of $0.3~r_s<r<10r_s$ or halo stars (right panel) in the radial range of $r_s<r<10r_s$. 

In both panels of Figure~\ref{fig:beta_dm_mc_glo}, we can see the dependence of $\beta$ on $M_{200}$ and $c_{200}$. At $c_{200}<10$, we can clearly see the monotonic trend that $\beta$ increases with the increase of $c_{200}$ in a fixed bin of $M_{200}$. Nevertheless, for the yellow and green curves with the smallest $M_{200}$, $\beta$ tends to decrease with the increase in $c_{200}$ at $c_{200}>\sim10$. For these few exceptional data points, we have checked their companions, and find many of them can have a massive companions within 4-5~Mpc. Moreover, many of them have peak mass happened in the past according to their mass assembly histories, and their $z=0$ mass already starts to drop compared with the peak mass. In our analysis, we only include those galaxy systems that are identified as central galaxies in TNG, but these halo systems close to a massive companion might have already undergo strong tidal effects. Their outer dark matter halos have already started being stripped by tides, leaving the central highly concentrated cores that are dominated by more isotropic orbits, hence causing such a decrease in $\beta$ with the increase in $c_{200}$. 

According to previous figures in this subsection, it is clear that for both dark matter particles and halo stars, their velocity anisotropy increases with the increase of $c_{200}$, at fixed $M_{200}$. A similar result has been found by Meng et al. (in preparation), and more detailed investigations and explanations on this trend will be available in Meng et al. (in preparation). Here we only make brief discussions about the possibilities. First of all, we think this is because, at fixed $M_{200}$, more concentrated dark matter halos can accrete and strip satellites/subhalos more efficiently. Satellites/subhalos are stripped earlier, during the earlier orbital stages when the motions are more radial, hence increasing the velocity anisotropy of accreted objects. Moreover, as will be discussed in more details by Meng et al. (in preparation), this is associated with the formation history of dark matter halos. For halos with larger concentrations, they form earlier. Though they have similar host halo mass at $z=0$, halos formed earlier have larger halo mass at higher redshifts, which are then expected to dominate the local environment at higher redshifts, thus resulting in more radial infalling of surrounding smaller satellite galaxies.

\subsection{Extensions to smaller mass ranges using TNG100 and TNG50}
\label{sec:ext}

So far, all the above results are based on TNG300 or TNG300-Dark. Due to the larger box six, we are able to have enough number of massive galaxy systems. However, the lowest resolution TNG300 simulation prevents us from going down to mass ranges smaller than $\log_{10}M_{200}/\msun\sim12.6$, below which the number of star particles is not enough to have realistic stellar halos. 

In this subsection, we move on to use the higher resolution TNG100 and TNG50 simulations to push our investigations down to smaller mass ranges. We also provide a fitting formula to the velocity anisotropy difference between halo stars and dark matter particles over a wider halo mass range, jointly using TNG300, TNG100 and TNG50. 

We first show the velocity anisotropy profiles of halo stars, dark matter particles and subhalos from TNG100 and TNG50 in Figure~\ref{fig:TNGall}. The legend shows the log mass ranges of $\log_{10}M_{200}/\msun$. Thanks to their higher resolutions, we can push down to $\log_{10}M_{200}/\msun\sim 11.4-11.8$, but due to the smaller simulation box size and less number of galaxy systems, the measurements are significantly noisier. Nevertheless, we can still see similar trends as in previous figures, that the outer velocity anisotropy profiles increase with the increase in $M_{200}$ for dark matter particles and subhalos. For halo stars, however, it is very difficult to see clear trends given the large errors. In the inner regions within $r_s$, the trends are also less clear to be seen given the large errors.

We summarize how $\beta$ of halo stars, dark matter particles and subhalos depends on $M_{200}$ in Figure~\ref{fig:beta_1-10rs}, using TNG300, TNG100 and TNG50 jointly to cover a wider mass range. We can clearly see that over the mass range of $11.4<\log_{10}M_{200}<14.8$, the velocity anisotropies for halo stars, dark matter particles and subhalos all tend to increase with the increase in host halo mass. The trend is the strongest for subhalos, and is less strong for dark mater particles. For halo stars, the trend is the weakest and can only be seen in the red solid curve based on TNG300. For the green and blue solid curves based on TNG100 and TNG50, perhaps due to the smaller box size, the measurements are noisy, which likely prevent us from clearly seeing the trend.  

Note the chosen radial range of halo stars, dark matter particles and subhalos in Figure~\ref{fig:beta_1-10rs} is $r_s<r<10~r_s$, so different from the varying inner radius cut adopted in previous figures. The same inner radius cut ensures more fair comparisons. For dark matter particles, the critical radius of the transition between inner and outer trends is in fact $r\sim 0.3~r_s$, but a cut at larger radius of $r>r_s$ would not affect the trend in the outer region. According to Figure~\ref{fig:beta_1-10rs}, we jointly fit a linear relation to $\beta_\mathrm{star}$ versus $\log_{10}M_{200}/\msun$, by using TNG30, TNG100 and TNG50 together. The best-fitting formula is $\beta_\mathrm{star}-\beta_\mathrm{DM}=(-0.028\pm 0.008)\log_{10}M_{200}/\msun + (0.690\pm0.010)$. More details are provided in the Appendix. 

\section{Discussions and conclusions}
\label{sec:concl}

In this paper, we perform detailed investigations on how do the velocity anisotropy parameters ($\beta$) of halo stars, dark matter particles and subhalos depend on the host halo properties, including the virial mass, $M_{200}$, and concentration, $c_{200}$. Our calculations are based on ex-situ star particles (we call as halo stars), dark matter particles (particles bound to surviving subhalos not included) and subhalos with more than 20 bound dark matter particles from the IllustrisTNG suites of simulations. 

We start our analysis by using the hydro-dynamical TNG300 simulation, by considering its large box size and thus enough number of massive galaxy systems. We find, over the mass range of $12.6<\log_{10}M_{200}/\msun<15.0$, the velocity anisotropy of halo stars, dark matter particles and subhalos all become more radial with the increase in their host halo mass, $M_{200}$, beyond a certain critical radius, and this trend becomes stronger with the increase in galactocentric radius of $r$. For halo stars, dark matter particles and subhalos, the critical radii are $r\sim r_s$, $r\sim0.3~r_s$ and $r\sim r_s$, respectively. Here $r_s$ is the scale radius of host dark matter halos in the NFW profile. There are not enough number of subhalos at small radii for robust conclusions.

We attribute such a monotonic trend at outer radii to the infalling orbits of satellite galaxies or subhalos. More massive dark matter halos dominate their surrounding local density field, resulting in more radial infalling of smaller satellite galaxies or subhalos \citep[e.g.][]{2020ApJ...905..177L}. Moreover, massive halos assemble late, and thus they are surrounded by a more active infall region with prominent radial motions, while low mass halos have nearly depleted their environment~\citep{FH21,Gao23}. The trend is weakened at smaller radius, because satellites/subhalos have lost memories of their initial infalling orbits, and satellites/subhalos with more radial orbits are depleted due to the strong tides in the center. 

On the contrary and in the more inner halo regions, the $\beta$ of both halo stars and dark matter particles become more isotropic with the increase of $M_{200}$. This is likely due to the dominance of in-situ baryons, that the in-situ galaxy is dominated by elliptical/spheroidal morphology, composed of more isotropic orbits. Though only ex-situ stars are used in our calculation, they could have exchanged energy and angular momentum through dynamical interaction with the in-situ particles, hence becoming more isotropic as well. 

We repeat our analysis with dark matter particles from the dark matter only (DMO) version of TNG300, i.e., TNG300-Dark, and identified a significantly different trend in the inner region, in terms that the velocity anisotropies show very weak dependence on $M_{200}$ for dark matter particles within $r=r_s$ and from TNG300-Dark. Moreover, dark matter particles from the hydro TNG300 are slightly more isotropic at $0.2~r_s<r<10~r_s$ and in halos with $\log_{10}M_{200}/\msun<13.8$ than TNG300-Dark. One possibility causing this is that the existence of baryons has caused subhalos to be more resistant to tidal effects, and thus their disruptions are delayed until getting to more central regions, resulting in flattened $\beta$ profiles. %Moreover, the underlying potential likely becomes more isotropic, due to the more isotropically distributed hot gas, for example, hence bringing in more isotropic orbits. 

We find that halo stars show the most radial $\beta$, which are larger than those of dark matter particles by approximately a constant beyond $r_s$, while the $\beta$ of dark matter particles are more radial than those of subhalos. We attribute this to the selection bias that halo stars were the most bound part of subhalos, and thus they more likely come from satellites/subhalos that have more radial orbits, hence passing close to the center and undergoing much stronger tidal strippings. Due to the selection bias, the dependence of $\beta$ on $M_{200}$ is the weakest for halo stars, and the strongest for subhalos, with dark matter particles in between. 

The fact that the $\beta$ profile of halo stars offsets above that of dark matter by approximately a constant is interesting, which suggests that if one can precisely measure the velocity anisotropy profile for observed halo stars, it can be used to further infer the $\beta$ profile for dark matter at all radius by adding this value. We provide a fitting formula to describe the difference between the velocity anisotropy of halo stars and that of dark matter particles at $r>r_s$ as $\beta_\mathrm{star}-\beta_\mathrm{DM}=(-0.028\pm 0.008)\log_{10}M_{200}/\msun + (0.690\pm0.010)$.

In addition to the dependence on $M_{200}$, we have also detected a monotonic trend that the $\beta$ of both halo stars and dark matter particles become more radial with the increase in the halo concentration, $c_{200}$, with fixed $M_{200}$. This is likely because more concentrated dark matter halos can accrete and strip satellites/subhalos more efficiently, hence increasing the velocity anisotropy. Moreover, this could be related to the assembly histories of galaxy/halo systems, that halos with higher $c_{200}$ assemble earlier, hence having larger masses at higher redshifts, resulting in more radial infalling of surrounding satellites/subhalos. 

In the end, we push our measurements down to smaller mass ranges of $\log_{10}M_{200}/\msun\sim11.4$ using the higher resolution TNG100 and TNG50 simulations. Due to their smaller box size, we fail to see prominent dependence of $\beta$ on $M_{200}$ for halo stars from TNG100 or TNG50, but we see similar trends for dark matter particles and subhalos at outer radii. Note halo stars show the weakest amount of increase with the increase of $M_{200}$ even in TNG300, than those of dark matter particles and subhalos, so the noise due to smaller samples could have dominates over the real signal in TNG100 and TNG50 for halo stars. 

Our detected trends of how $\beta$ changes with the host halo mass and concentration can be used as a theoretical reference for studies of tracer velocity anisotropies of our Milky Way and extra-galactic systems ranging from massive ellipticals to sub Milky Way mass galaxies. It can also be used to set up a broad prior upon constraining Milky Way halo mass using dynamical tracers, when the velocity anisotropies of the tracer population are not accurately known due to missing dimension of data or large observational errors. 

\acknowledgments

This work is supported by NSFC (12273021, 12022307), the National Key R\&D Program of
China (2023YFA1605600, 2023YFA1605601, 2023YFA1607800, 2023YFA1607801), the China Manned Space (CSST) 
Project with No. CMS-CSST-2021-A02 and No. CMS-CSST-2021-A03, the National Key Basic Research and Development Program of China (No.~2018YFA0404504), 111 project (No.~B20019) and Shanghai Natural Science Foundation (No. 19ZR1466800). We thank the sponsorship from Yangyang Development Fund. Z-L acknowledges the funding from the European Unions Horizon 2020 research and innovation programme under the Marie Skodowska-Curie grant 101109759 (“CuspCore”). The computation of this work is carried out on the \textsc{Gravity} supercomputer at the Department of Astronomy, Shanghai Jiao Tong University. The first author of this paper accomplished calculations and figure plotting of this paper during his master studies. The paper writing is mostly finished by the corresponding author. 

\appendix

\section{A fitting formula to the velocity anisotropy difference between halo stars and dark matter}

\begin{figure}
\centering
\includegraphics[width=0.8\linewidth]{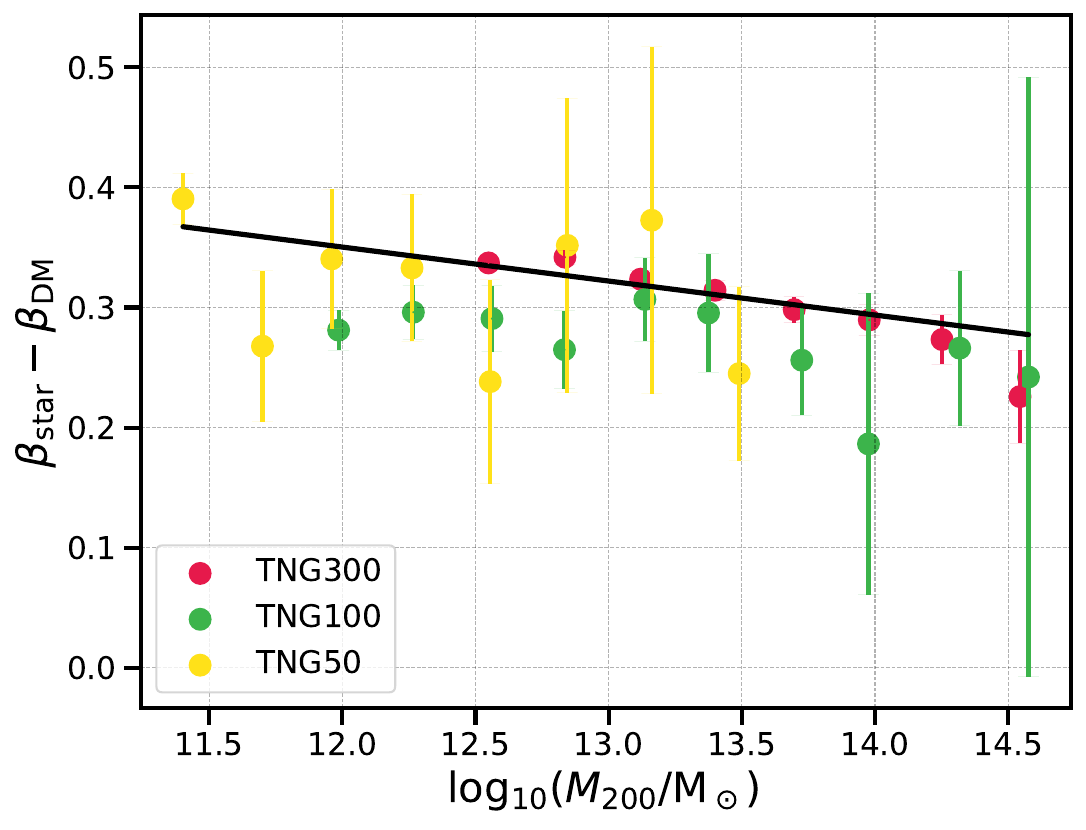}
\caption{A joint fit to the velocity anisotropy difference between that of halo stars and of dark matter particles, using results from TNG300, TNG100 and TNG50 together. Here the velocity anisotropies of halo stars and dark matter particles are calculated based on particles with galactocentric radii of $r_s<r<10~r_s$.}
\label{fig:fit}
\end{figure}

In the main text of this paper, we noticed that the difference between the velocity anisotropy of halo stars and dark matter particles is approximately a constant beyond galactocentric radius of $r>r_s$, and also shows a weak halo mass dependence. Using the velocity anisotropy difference between that of halo stars and of dark matter particles measured from TNG300, TNG100 and TNG50, which covers a wide halo mass range, we perform a joint fitting between the relation of $\beta$ and $\log_{10}M_{200}/\msun$. Here the velocity anisotropy is calculated using halo stars and dark matter particles with galactocentric radius of $r_s<r<10~r_s$. The velocity anisotropy difference between halo stars and dark matter particles are shown by symbols with errorbars in Figure~\ref{fig:fit}, and the black solid line is the best-fitting result of $\beta_\mathrm{star}-\beta_\mathrm{DM}=(-0.028\pm 0.008)\log_{10}M_{200}/\msun + (0.690\pm0.010)$. Note the symbols are based on the difference between the halo star and dark matter measurements in Figure~\ref{fig:beta_1-10rs} of the main text. 

\clearpage

\bibliography{master}

\end{document}